\documentclass{aa}
\usepackage{graphicx,color}
\usepackage{natbib}
\usepackage{url}
\usepackage{amsmath}
\usepackage{longtable}

\bibpunct{(}{)}{;}{a}{}{,} 
\newcommand{\ulyss}{ULySS}

\begin{document}
 \title{Coud\'e-feed stellar spectral library - atmospheric parameters}

  \author{ Yue Wu,
 \inst{1,2}
 Harinder P. Singh,
 \inst{3,1}
 Philippe Prugniel,
          \inst{1}
 Ranjan Gupta
 \inst{4,1}
 \and
 Mina Koleva
 \inst{5,1}
        }

   \offprints{Ph. Prugniel}

   \institute{Universit\'e de Lyon, Universit\'e Lyon 1, Villeurbanne, F-69622, France;
              CRAL, Observatoire de Lyon, CNRS UMR~5574, 69561 Saint-Genis Laval, France\\
              \email{prugniel@obs.univ-lyon1.fr}
              \and
National Astronomical Observatories, Chinese Academy of Sciences, 20A Datun Road, Chaoyang District, Beijing 100012, China;
              Key Laboratory of Optical Astronomy, NAOC, Chinese Academy of Sciences;
              Graduate University of the Chinese Academy of Sciences, 19A Yuquan Road, Shijingshan District, Beijing, 100049, China\\
              \email{wuyue@lamost.org}
              \and
              Department of Physics and Astrophysics, University of Delhi, India\\
              \email{hpsingh@physics.du.ac.in}
              \and
              IUCAA, Post Bag 4, Ganeshkhind, Pune 411007, India\\
              \email{rag@iucaa.ernet.in}
              \and
              Instituto de Astrof\'{\i}sica de Canarias, La Laguna, E-38200 Tenerife, Spain; Departamento de Astrof\'{\i}sica, Universidad de La Laguna, E-38205 La
Laguna, Tenerife, Spain\\
              \email{koleva@iac.es}
 }

   \date{Received ; accepted }


  \abstract
  {
Empirical libraries of stellar spectra play an important role in
different fields. For example, they are used as reference for the
automatic determination of atmospheric parameters, or for building
synthetic stellar populations to study galaxies. The CFLIB
(Coud\'e-feed library, Indo-US) database is at present one of the
most complete libraries, in terms of its coverage of the atmospheric
parameters space ($T_{\rm{eff}}$, log~$g$ and [Fe/H]) and wavelength
coverage 3460 - 9464 \AA{} at a resolution of $\sim$1 \AA{} FWHM.
Although the atmospheric parameters of most of the stars were
determined from detailed analyses of high-resolution spectra, for
nearly 300 of the 1273 stars of the library at least one of the
three parameters is missing. For the others, the measurements,
compiled from the literature, are inhomogeneous. }
  {In this paper, we re-determine the atmospheric parameters,
directly using the CFLIB spectra, and compare them to the previous studies.
}
    {
We use the \ulyss{} program to derive the atmospheric parameters, using the
ELODIE library as a reference.
     }
     {
Based on comparisons with several previous studies we conclude that
our determinations are unbiased. For the 958 F,G, and K type stars
the precision on $T_{\rm{eff}}$, log~$g$, and [Fe/H] is respectively
43~K, 0.13~dex and 0.05~dex. For the 53 M stars they are 82~K,
0.22~dex and 0.28~dex. And for the 260 OBA type stars the relative
precision on $T_{\rm{eff}}$ is 5.1\%, and on log~$g$, and [Fe/H] the
precision is respectively 0.19~dex and 0.16~dex. These parameters
will be used to re-calibrate the CFLIB fluxes and to produce
synthetic spectra of stellar populations.}
   {}
 \authorrunning{Wu et al.}
   \titlerunning{Stellar atmospheric parameters for CFLIB}

  \maketitle
%

\section{Introduction}

Spectral libraries with a good coverage in $T_{\rm{eff}}$, log~$g$,
and  [Fe/H]  at moderate spectral resolution (1-2 ~\AA~FWHM) are
essential in several areas of astronomy. Two important applications
are the spectral synthesis of the stellar population of galaxies
\citep[e. g.][]{bc03,phr,vaz10} and the automated classification and
determination of the stellar atmospheric parameters \citep[e.
g.][]{Katz98}. The latter application is a key step in the analysis
of large spectroscopic surveys, such as RAVE \citep{stein06},
SDSS/SEGUE, APOGEE and SEGUE-2 \citep{rock09}, the Guoshoujing
Telescope\footnote{formerly named LAMOST,
\url{http://www.lamost.org}} survey \citep{zhao06} or GAIA
\citep{perr01}.
Both the increased output of the modern instruments and the better
quality of the spectra require improvements of the spectral
libraries and models.

The first large library (684 stars) at the intermediate spectral
resolution of $\sim$ 2 ~\AA{} (FWHM) was published by \citet{Jones}.
It covered only two small windows of the optical range with an
incomplete sampling of the atmospheric parameters. It was followed
by more complete libraries improving the wavelength and parameter
coverage. The most significant ones at present are the ELODIE
\citep{PS01}, CFLIB \citep{cflib} and MILES \citep{miles}.

The ELODIE library \citep{PS01,elo31} has a compilation of 1962
spectra of 1070 stars observed at a resolution $R \sim 42000$ in the
range 3900-6800~\AA. The library is also available at a resolution
of 0.55 \AA{} ($R \sim 10000$) for the population synthesis in
PegaseHR \citep{phr}. The main limitation of this library is its
restricted wavelength coverage. The CFLIB (also called Indo-US
library) has a wide wavelength coverage (3460-9464~\AA) at a lower
resolution ($R \sim 5000$), its most important limitation is its
approximate flux calibration. The youngest library in the optical
range is MILES \citep{miles}. With a wavelength coverage from 3525 -
7500~\AA, it is limited in its coverage in the red compared to
CFLIB. Its resolution is only 2.3~\AA{} (FWHM, $R \sim 2200$), but
its flux calibration is undoubtedly more precise and is used for
population synthesis \citep{vaz10}.

The reliability and accuracy of the atmospheric parameters of the
stars of these libraries has direct consequences on their usage, for
example in stellar populations models
\citep{pru07,kol07,per09,chen10}. While most of the stars of CFLIB
have atmospheric parameters compiled from the literature and in
general determined from high-dispersion spectra, there is a
substantial fraction ($\sim 25\%$) of stars with either one or more
parameters unavailable. Because the flux calibration was done by
fitting each observation to a spectral energy distribution with a
close match in spectral type from the \cite{pickles1} library, the
stars with poorly known atmospheric parameters were also
inaccurately calibrated. The present paper intends to re-measure
these parameters homogeneously and is therefore a step toward an
improvement of the flux calibration.

Various methods have been developed in the past to estimate stellar
atmospheric parameters from stellar spectra in an automatic and
reliable manner. One of the often used technique involves finding
the minimum distance between observed spectra and grids of synthetic
or observed spectra. The program TGMET, developed by \citet{Katz98}
and improved by \citet{Soub00, Soub03}, illustrates this approach.
It achieves an internal accuracy of 86~K, 0.28~dex and 0.16~dex for
$T_{\rm{eff}}$, log~$g$, and [Fe/H] respectively for a target F, G,
or K star with signal-to-noise ratio S/N~=~100 and 102~K, 0.29~dex
and 0.17~dex at S/N~=~10.

\citet{fior} derived atmospheric parameters from observed
medium-resolution stellar spectra using non-linear regression models
trained either on pre-classified observed data or synthetic stellar
spectra. For the SDSS/SEGUE spectra, they reached an accuracy on the
order of 150~K in the estimation of $T_{\rm{eff}}$, 0.36 dex in
log~$g$, and 0.19 dex in [Fe/H]. Other similar efforts include
\cite{BJ97}, \cite{PS01},  \cite{Snider01}, \cite{Willem05},
\cite{RB06}, \cite{Shkedy07}, \cite{lal08}, \cite{lefever09},
\cite{jnzh09} and \cite{jofre10}.

The method employed in this work uses the
\ulyss\footnote{\url{http://ulyss.univ-lyon1.fr}} package and
consists of minimizing a $\chi^2$ between an observed
 spectrum and a model spectrum. This model is adjusted at the same
resolution and sampling as the observation, and the fit is performed
in the pixel space. The method determines all free parameters in a
single fit in order to handle properly the degeneracy between the
temperature and the metalicity.

The details of the method and the process of extraction are
presented in the next section. In Sect. 3 we test the quality of
parameter extraction by comparing  with other determinations in the
literature. In Sect. 4, we give the results and conclusions of our
study.



\section{Analysis method}

The \ulyss{} package fits a spectrum against a non-linear model. The
package has flexibility allowing one to use it for various types of
tasks. For example, in \citet{koleva09} it was used to retrieve the
star-formation history of galaxies and in \citet{wu10} to analyze
stellar spectra. In the present work, we will use the so-called TGM
component of the package to fit the CFLIB spectra against stellar
spectral models based on a new version of the ELODIE library
(version 3.2).  In this section, we describe this new library and
the fitting method.

\subsection{ELODIE 3.2}
\label{sect:elo32}

The ELODIE library is based on echelle spectra taken with the eponym
spectrograph attached to the 1.93 m telescope of Observatoire de
Haute-Provence. The data used to prepare the library were already
processed by the data-reduction pipeline run at the telescope. They
can be retrieved from the observatory archive \citep{moultaka04}.

The first version of the library \citep{PS01} contained 908 spectra
for 709 stars, and the spectra were provided in the wavelength range
4100-6800~\AA. Two upgrades were released in \citet[version
3.0]{PS04} and \citet[version 3.1]{elo31}. The version 3 series
rests on a new and larger selection of spectra from the whole
archive of the instrument. There are 1962 spectra for 1070 stars
within wavelength range 3900-6800~\AA{} with a larger coverage of
atmospheric parameter space.

The three lines of improvement that motivated the successive
releases were (i) the data-reduction (in particular the correction
of the diffuse light and the extension of the wavelength range to
3900-6800~\AA), (ii) the determination of the atmospheric parameters
and Galactic extinctions and (iii) the interpolator. This last point
is particularly important for the present study because it is the
function that returns a spectrum for a given set of atmospheric
parameters by making an interpolation over the whole library
\citep{prugniel08}.

Since the release of version 3.1, the ELODIE library has thus been
continuously improved, and for the present work we use the version
3.2 that will also be described separately (Prugniel et al. in
preparation). The main motivation to use this new version here is a
modification of the interpolator, which now takes the natural
broadening of the spectra (due to rotation) as an additional
parameter.

The interpolator consists of polynomial expansions of each
wavelength element in powers of log(T$_{\rm{eff}}$), log~$g$, [Fe/H]
and $f(\sigma)$. The last term is a function of the rotational
broadening parameterized by $\sigma$, the standard deviation of a
Gaussian. We used $f(\sigma) = 1 - 1/\sqrt{1+\sigma^2}$, with
$\sigma$ in pixels, because it is approximately proportional to the
change of depth of the lines caused by the broadening. The Gaussian
broadening of each spectrum was determined with \ulyss{} using a
previous version of the model, and is, therefore, relative to the
'mean' rotational width for a given spectral type.

Three sets of polynomials are defined for three temperature ranges
(roughly matching the OBA, FGK, and M types) with considerable
overlap between them where they are linearly interpolated. For the
FGK and M polynomials, 26 terms are used and for the OBA
polynomials, 19 terms are used. The terms of the polynomials were
chosen to be almost orthogonal, for ease of understanding their
contribution and to avoid contributions of large amplitudes that
cancel each other and result in a loss of precision. The inclusion
of new terms was made gradually, testing those whose contribution
appears to be the most important, as measured by the decrease of the
total squared residuals between the original spectra and the
interpolated ones. Another argument that guided the choice of the
developments was the minimization of the biases between the
atmospheric parameters compiled from the literature (absolute
calibration), and the internal values obtained by inverting the
interpolator. These biases were probed in different regions of the
space of parameters. The final choice of the developments and the
limit tuning of the three temperature ranges were based on
intuitions and tests.

The coefficients of the polynomials were fitted on a sub-set of the
whole library after excluding peculiar stars (these stars are listed
in the table of the ELODIE 3.1 release; the corresponding list for
the present version is almost unchanged and will be given in
Prugniel et al. in preparation). A first interpolator is based on
parameters compiled from the literature (that we call {\it absolute}
parameters). For the stars without spectroscopic determinations of
the parameters, we used photometric calibrations, or a calibration
from the spectral classifications, which are obviously less accurate
than the spectroscopic ones. The quality of this {\it absolute}
interpolator is limited by the inhomogeneity and inaccuracy of the
atmospheric parameters. Therefore, we fitted the observed spectra
against the interpolated ones to derive revised parameters for those
stars that were photometrically calibrated, and we adopted the
average between the original and new parameters. We computed a
second interpolator, based on these improved parameters, and used it
in an iteration to measure homogeneously the atmospheric parameters
for the complete library. These self-calibrated {\it internal}
parameters are finally used to compute an {\it internal}
interpolator, which is used for the stellar population synthesis
with PegaseHR. The {\it internal} interpolator is more precise in
the sense that the residuals between the original observations and
the interpolated spectra are reduced, but it may be affected by
biases owing to the imperfect modeling by polynomials.

In previous versions, the inversion of the interpolator restored
biased atmospheric parameters for the intermediate temperature
stars. For example, for the 13 observations of the Sun, the mean
temperature obtained in version 3.1 is 5726~$\pm$~11~K, compared to
the reference temperature of 5777~K (i.e. the one adopted by
\citealt{elo31} from \citealt{cayrel96}). A careful examination
revealed that this negative bias was balanced with a positive bias
for the fast rotating stars. The effect of the rotation reduces the
depth of the lines, which then can be fitted to a hotter
temperature. Surprisingly this degeneracy links the rotation more to
the temperature than to the metalicity\footnote{By contrast, in
stellar populations the broadening owing to the velocity dispersion
is clearly degenerated with the metalicity \citep{koleva08}.}. This
analysis lead us to introduce rotation terms in to the polynomial
developments. With the new version, the temperature of the Sun is
5760~$\pm$~14~K. Even if we exclude three observations from the same
date (1999/12/22) which depart significantly from the 10 others, the
mean temperature is 5766~$\pm$~9~K. For the same set of
observations, log~$g$ and [Fe/H] are respectively 4.33~$\pm$~0.03
and -0.03~$\pm$~0.02 dex compared to the reference values of
4.44~log($g$/cm.s$^2$) and 0.0 dex respectively. For some reason
there is apparently a small but significant bias on the surface
gravity of the Sun, but the consistency for the temperature is
extremely good. The comparison between the absolute and internal
determination of the atmospheric parameters in various small regions
of the parameter space did not reveal noticeable biases.

Another improvement in the new version concerns the ability of the
interpolator to model the spectra in sparsely populated regions of
the parameter space, and even to extrapolate out of the range of
parameters of the library.

This has been achieved by supplementing the observed spectra with
'semi-empirical' ones. We computed these semi-empirical spectra by
adding the differential effect predicted in the theoretical
libraries \citep{coelho05,martins05} between two points of the
parameter space to the interpolated spectrum in a reference point.
We can write it as

\begin{eqnarray*}
S_{s}(\:T_{1},\,\log\!g_{1},\,[Fe/H]_{1}) & = & S_{i}(\:T_{0},\,\log\!g_{0},\,[Fe/H]_{0}) + \\
& &\big[ \:S_{t}(\:T_{1},\,\log\!g_{1},\,[Fe/H]_{1}) - \\
& & S_{t}(\:T_{0},\,\log\!g_{0},\,[Fe/H]_{0}) \:\big].
\end{eqnarray*}

While ($T_{0}$, log~$g_{0}$, $[Fe/H]_{0}$) is a reference point in
the parameter space where the interpolator predicts a reliable
spectrum, and ($T_{1}$, log~$g_{1}$, $[Fe/H]_{1}$) is a point
located outside of the region populated by the library stars,
$S_{i}$ is a spectrum predicted by the interpolator (previous
version) and $S_{t}$ are spectra from the theoretical library.
$S_{s}$ is the semi-empirical spectrum used to complete the library.
This method guarantees the continuity between the observed and
theoretical libraries and provides better approximations of the
spectra than the pure theoretical spectra, which are devoid of some
lines. We used these semi-empirical spectra to constrain the
extrapolations toward very hot and cool stars and toward the low
metalicities for hot stars. The spectra from \citet{coelho05} were
used to extend the interpolator for the warm stars to low metalicity
(down to [Fe/H]~=~-2.5~dex), and those from \citet{martins05}, to
extend to high and low temperatures. Note that despite a more
sophisticated model (Phoenix, \citealp{hauschildt96}), these latter
models do not match the observations for the cool stars,
$T_{\rm{eff}}<$3500~K, in the spectral region 6300 to 6500~\AA. This
is certainly because of a wrong line list for the molecular
absorption (Martins, private communication).

The modeling of the spectra with polynomials is similar to the
fitting functions for the Lick spectroscopic indices
\citep{gorgas,worthey94}. As an alternative to this global
interpolation, \citet{vazdekis03} developed a Gaussian kernel
smoothing, which may be less sensitive to biases, but is more
sensitive to the errors on the atmospheric parameters and to the
particularities of the stars in the scarcely populated regions of
the space of parameters. In these regions, only a few stars are used
for the local interpolation and the resulting spectra are therefore
more subject to the cosmic variance, while this effect is smoothed
for a global interpolation.

We computed interpolators both for the continuum-normalized and for
the flux-calibrated versions of the library. The residuals between
the observed and interpolated spectra for continuum-normalized
version are lower than for the flux-calibrated version because of
the additional uncertainties of the flux-calibration and correction
of the Galactic extinction. In this work, we adopted the
flux-calibrated version to analyze the flux-calibrated CFLIB
observations.

\subsection{Fitting method}

The function described above is interfaced with \ulyss{} to fit the
observed spectrum with an interpolated spectrum convolved with a
Gaussian and multiplied by a polynomial. The order of this
polynomial is chosen to absorb any flux calibration systematics or
effects of extinction. The minimization problem is written as

\vspace {0.5cm} $\rm Obs(\lambda) = P_{n}(\lambda) \times
[\,\rm{TGM}(\,T_{\rm{eff}},log~$g$,[Fe/H],\lambda) \otimes
G(\,v_{sys},\sigma)]$, \vspace {0.5cm}

where $\rm Obs(\lambda)$ is the observed spectrum, $\rm
P_{n}(\lambda)$ a Legendre polynomial of degree n, and $\rm
G(v_{sys},\sigma)$ is a Gaussian broadening function characterized
by the systemic velocity $v_{sys}$, and the dispersion $\sigma$. The
free parameters of the minimization procedure are the three
parameters of the TGM function: $T_{\rm{eff}}$, log~$g$ and [Fe/H],
the two parameters of the Gaussian: $v_{sys}$, $\sigma$ and the
coefficients of $P_{n}$. $v_{sys}$ absorbs the imprecision of the
cataloged radial velocity of the stars that were used to reduce them
in the rest frame; $\sigma$ encompasses both the instrumental
broadening and the effect of the rotation.

\subsection{Consistency tests}
 \label{sec.consistency}

As a consistency test, we determined the atmospheric parameters of
the ELODIE library stars using \ulyss{}. We compared the
continuum-normalized spectra to the continuum-normalized
interpolator 
based on the absolute parameters (i.e. compiled from the
literature). We fitted the three atmospheric parameters and a
broadening function, which accounts for rotation, with a rejection
of the outliers.

Because the {\it internal} atmospheric parameters determined in the
ELODIE library were also obtained from an inversion of this
interpolator, we expect \ulyss{} to return very similar results.
Though the two minimization approaches are different (an ad-hoc
downhill method designed to avoid known local minima in the case of
ELODIE), both methods minimize the squared departures between the
model and the observation computed on each wavelength bin. Still,
the two analyses differ by significant details. In particular, while
ULySS matches the rotation using a Gaussian convolution, in ELODIE
it is included as specific terms in the polynomial developments of
the interpolator (see Sect.~\ref{sect:elo32}).

For this test, we used only  ELODIE observations with reliable
absolute parameters. The number of comparison spectra are 293 for O,
B, and A, 1415 for F, G, and K and 26 for M types. The biases and
dispersions with respect to the {\it internal} determinations of the
ELODIE library are shown in the left part of Table~\ref{tb.test}.
The first line gives the mean formal fitting errors. The second
gives the statistics using all the measurements, and the third, the
statistics after excluding the values departing by more than
3$\sigma$ from the mean.

The biases between the two series of measurements are not
significant and the dispersions are small compared to the usual
precision on this type of determinations. After checking the most
deviating stars, we conclude that the main source of discrepancy
comes from different modeling of the rotational broadening. The
convolution method of \ulyss{} is better, and the ELODIE
interpolator should be improved. However, as this is well within the
final expected errors, we will proceed with the current version.

\begin{table*}
\caption{\label{tb.test}Consistency tests with the continuum-normalized interpolator
and with the flux-calibrated interpolator.}
\begin{tabular}{l|cc|cc|cc||l|cc|cc|cc}
\hline\hline
Sp.Type\tablefootmark{a}
& \multicolumn{2}{c|}{$\Delta$$T_{eff}$} & \multicolumn{2}{c|}{$\Delta$log$g$($cm\,s^{-2}$)} & \multicolumn{2}{c||}{$\Delta$[Fe/H](dex)} & Sp.Type\tablefootmark{a} & \multicolumn{2}{c|}{$\Delta$$T_{eff}$} & \multicolumn{2}{c|}{$\Delta$log$g$($cm\,s^{-2}$)} & \multicolumn{2}{c}{$\Delta$[Fe/H](dex)} \\
          & mean & rms & mean & rms & mean & rms & & mean & rms & mean & rms & mean & rms \\
\hline
OBA      &         & 0.17~\% &        & 0.008 &       & 0.006 & OBA      &        &  0.17~\% &        &  0.007 &       & 0.006 \\
9/293    &  0.21~\% & 3.1~\% & -0.022 & 0.129 & 0.029 & 0.128 & 13/293   & 0.39~\%&  4.3~\%  & -0.027 &  0.159 & 0.043 & 0.119 \\
         &  0.18~\% & 2.2~\% & -0.016 & 0.096 & 0.029 & 0.125 &          & 0.34~\%&  3.3~\%  & -0.027 &  0.129 & 0.042 & 0.116 \\
         &          &        &        &       &       &       &          &        &          &        &        &       &       \\
FGK      &          &  3.7~K &        & 0.006 &       & 0.003 & FGK      &        &  3.7~K   &        &  0.006 &       & 0.003 \\
69/1415  &  2.3~K   & 38.3~K & -0.022 & 0.071 & 0.023 & 0.065 & 37/1415  &-1.9~K  &  44.4~K  & -0.028 &  0.086 & 0.023 & 0.069 \\
         &  0.2~K   & 23.0~K & -0.021 & 0.067 & 0.019 & 0.055 &          &-1.1~K  &  29.5~K  & -0.025 &  0.079 & 0.023 & 0.063 \\
         &          &        &        &       &       &       &          &        &          &        &        &       &       \\
M        &          & 1.1~K  &        & 0.006 &       & 0.003 & M        &        &  1.1~K   &        &  0.006 &       & 0.003 \\
3/26     & -2.3~K   & 17.6~K &  0.005 & 0.181 & 0.003 & 0.051 & 2/26     &-2.9~K  & 17.8~K   & -0.005 &  0.227 & 0.005 & 0.054 \\
         & -0.5~K   & 11.8~K &  0.021 & 0.165 & 0.009 & 0.043 &          &-2.4~K  & 14.0~K   & -0.004 &  0.236 & 0.011 & 0.049 \\
\hline
\end{tabular}
\tablefoot{ The left panel presents the parameter statistics for the
continuum-normalized interpolator, and the right one for the
flux-calibrated interpolator. For each stellar type group, the first
line gives the formal fitting errors. The second gives the raw
statistics, and the third, the statistics after rejecting the 3
$\sigma$ $T_{\rm{eff}}$ outliers.
The comparisons are the \ulyss{} values minus the ELODIE values. \\
\tablefoottext{a}{The second line gives the total number of spectra used for the
comparison and the number of clipped $T_{\rm{eff}}$ outliers.}
}
\end{table*}

A second test consisted in analyzing the flux-calibrated ELODIE
spectra with \ulyss{} using the flux-calibrated interpolator and the
corresponding biases, and the dispersions are shown in
Table~\ref{tb.test} (right side).

The residuals of the comparisons between the two sets of atmospheric
parameters are only a little higher in this second test. This was
expected because this time the measurements differ not only by the
fitting method, but also by the models. We can a priori estimate
that the flux-calibrated interpolator is not as accurate as the
continuum-normalized one, since it also {\it averages} the flux
calibration errors (which are significant in ELODIE). We can also
assume that the flux-calibrated interpolator is less precise for the
same number of terms in the developments, since the spectra are more
complex (they include the thermal component).

A precision better than 40 K for a F, G and K star is fully
satisfactory for the present purpose, and we will not investigate
solutions to improve the flux-calibrated interpolator.

\subsection{Relative line-spread function}

In principle, an initial step prior to the analysis is to adapt the
resolution of the model to the one of the observation, this is
called line-spread function (LSF) injection. This is in general more
complex than a convolution, because the relative resolution between
the observation and the model may vary with wavelength (see the
detailed description in \citealt{ulyss}). However, as CFLIB was made
over a period of eight years, with slightly different setups, the
resolution also changes throughout the library. This would require a
separate analysis of the resolution for each spectrum. This can be
done with \ulyss{}, but after testing several cases for various type
of stars, we found that this was not producing significantly
different results than if we simply included a simple Gaussian
broadening in the fitted model. This broadening also matches the
rotational broadening of the fitted star.

\begin{figure}
\includegraphics{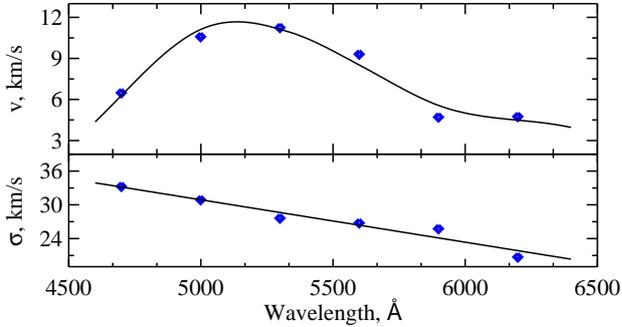}
\caption {
Relative line-spread function of CFLIB with respect to ELODIE for
HD\,4307. The blue diamonds are the values
derived using {\sc uly\_lsf}, and the black lines are
smoothed functions.}
\label{fig.lsf}
\end{figure}

\begin{figure*}
\includegraphics{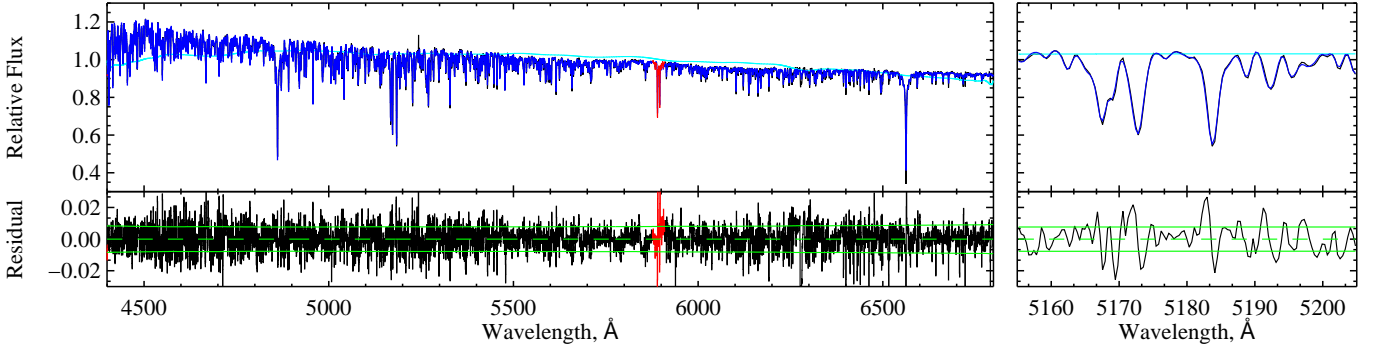}
\caption {Fit of the CFLIB observation HD\,4307 (G2V) with a TGM
component. The top panel shows the spectrum(in black) and the best
fit in blue (both are almost superimposed and the black line can be
seen only when zooming on the figure), the light blue is the
multiplicative polynomial. In red we plot the flagged and masked NaD
telluric lines that were not well calibrated in ELODIE library. The
residuals are plotted in the bottom panels. The continuous green
lines mark the 1-$\sigma$ deviation, and the dashed line is the
zero-axis. The right side shows a small wavelength region around
Mg$_b$. The order of the multiplicative polynomial is n=70.}
 \label{fiteg}
\end{figure*}

Figure~\ref{fig.lsf} shows the relative LSF between the ELODIE
library and a star from CFLIB (chosen randomly). It was evaluated
with wavelength intervals of 600 \AA{} spaced by 300 \AA{}. The
variation of the instrumental velocity dispersion ($\sigma_{ins}$)
with the wavelength is significant: from 21 to 34 $\rm km\,s^{-1}$.
For this star, the atmospheric parameters without LSF injection are
$T_{\rm{eff}}$=5803 K, log~$g$=4.01 dex, [Fe/H]=-0.26 dex. With LSF
injection they are $T_{\rm{eff}}$~=~5787~K, log~$g$~=~3.98~dex,
[Fe/H]=-0.25~dex. Figure~\ref{fiteg} shows the corresponding fit and
illustrates the general quality. We checked the same for 13 other
randomly selected CFLIB observations, which represent seven
different stellar spectral types, and concluded that the differences
are within the uncertainties.

\subsection{Error estimation}
 \label{sec.err}

In principle, to compute the formal fitting errors we need to know
the random errors on each wavelength element. Unfortunately, this
information is not provided with the reduced CFLIB spectra.
Therefore, we determined an upper limit to these internal errors by
assuming that the residuals are entirely due to the noise (i. e. the
fit is perfect and there are no residuals of physical origin). In
this case, the reduced $\chi^2$ is by definition equal to unity. To
make this determination, we assumed that the noise is the same on
all the wavelength elements. We performed the fit with an arbitrary
value of S/N and rescaled the errors returned by \ulyss{} by
multiplying them by $\sqrt{\chi^2}$. We verified that the
implementation is correct by checking that the errors do not depend
on the over-sampling degree of the data.

The mean errors for the fits of the ELODIE spectra (see
Table~\ref{tb.test}) are one order of magnitude smaller than the
dispersion between the inversion with \ulyss{} and the measurements
in the ELODIE library. This difference is certainly caused by the
correlation between the atmospheric parameters, in particular
between the temperature and the metalicity, as illustrated in
\citet[their fig. 7]{ulyss} and to the correlation between other
characteristics of the atmosphere (some other tests indicate that
the data are not over-fitted; see Sect.~\ref{sec.multip}). In any
case, the internal errors are likely proportional to the formal
fitting errors, and we estimate them by a simple rescaling to the
dispersions reported in the right panel of Table~\ref{tb.test},
according to the spectral type.

\subsection{Order of the multiplicative polynomial}
\label{sec.multip}

\begin{figure}
\includegraphics[width=9cm]{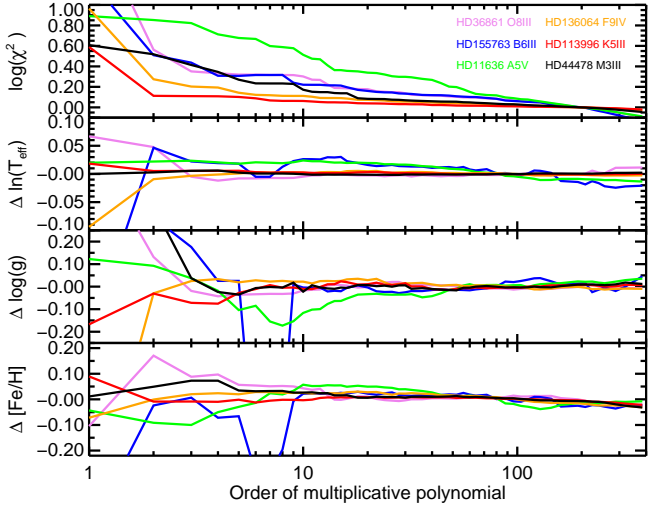}
\caption {Evolution of the stellar atmospheric parameter fit results
(log($\chi^2$), log($T_{\rm{eff}}$), log~$g$, and [Fe/H]) with
increasing Legendre polynomial degree.}
 \label{fig.poly}
\end{figure}

To determine the optimal order of the multiplicative polynomial,
$n$, we proceeded as suggested in \citep{ulyss}. We selected some
stars that are representative of all main spectral types and
analyzed them with a series of values of $n$ to find the point where
the solutions become independent of $n$. Figure~\ref{fig.poly},
compared with the similar graph presented in \citet{ulyss} shows the
improvement brought about by the new version of the interpolator:
the plateau is reached for a lower $n$ and the solutions are more
stable. In this work, we adopt a polynomial degree $n$ = 70.

\subsection{Analysis strategy}

\begin{figure}
\includegraphics[width=9cm]{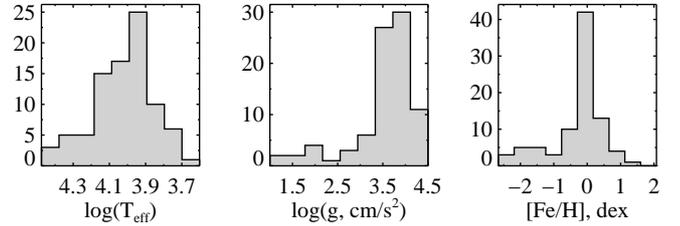}
\caption {Distribution of the 260 O, B, and A CFLIB stars with
already partially published atmospheric parameters, actually
including 87 $T_{\rm{eff}}$, 86 log~$g$ and 86 [Fe/H]. Note that the
ordinates of the three panels are not on the same scale.}
\label{fig.hist1}
\end{figure}

\begin{figure}
\includegraphics[width=9cm]{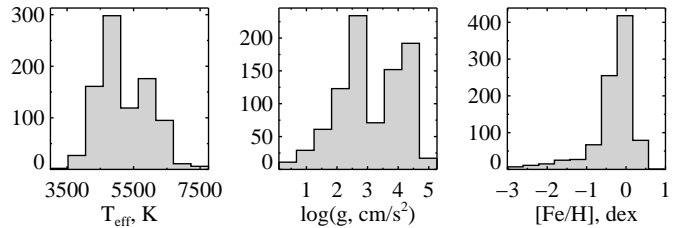}
\caption {Distribution of the 958 F, G, and K CFLIB stars with
already partially published atmospheric parameters, actually
including 895 $T_{\rm eff}$, 890 log~$g$ and 905 [Fe/H]. Note that
the ordinates of the three panels are not on the same scale.}
\label{fig.hist2}
\end{figure}

\begin{figure}
\includegraphics[width=9cm]{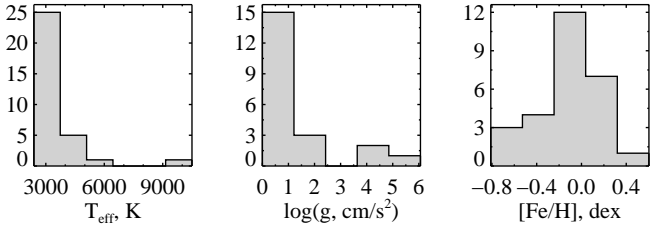}
\caption {Distribution of the 53 M CFLIB stars with already
partially published atmospheric parameters, actually including 32
$T_{\rm{eff}}$, 21 log~$g$ and 27 [Fe/H]. Note that the ordinates of
the three panels are not on the same scale.} \label{fig.hist3}
\end{figure}

We classified the stars into three groups, 260 O, B \& A (hereafter
OBA) , 958 F, G \& K (hereafter FGK) and  53 M type stars. The
distributions of the atmospheric parameters compiled in the CFLIB
paper are shown in Figs.~\ref{fig.hist1},~\ref{fig.hist2} and
~\ref{fig.hist3} for the OBA, FGK and M types respectively.

\begin{figure*}
\centering
\includegraphics{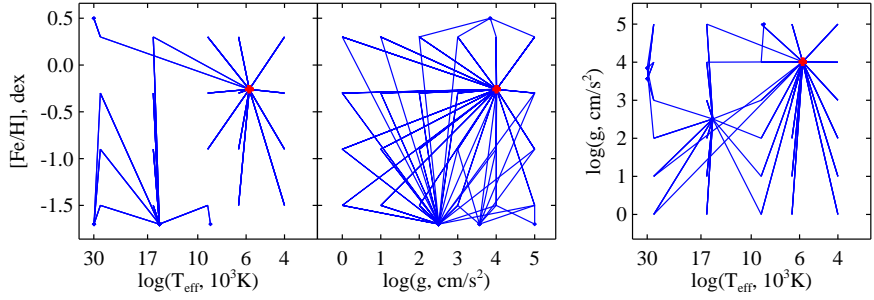}
\caption {Convergence maps on different projections of the
parameters space for the CFLIB star HD\,4307.
Each vector connects a guess with the correspondng solution.
The red dots show the location of absolute minimum in the three projections.
}
 \label{fig.cvg}
\end{figure*}

Because ULySS performs a {\it local} minimization, we studied the
structure of the parameter space to understand where local minima
may actually occur and trap the solution. We did this with
convergence maps for several stars that spanned the range of the
parameters. These tests (see an example in Fig.~\ref{fig.cvg})
consisted in fitting the spectra from a wide grid of guesses to
identify the convergence regions. According to the results we chose
the following grids of guesses for the three temperature groups:

\begin{itemize}
 \item O, B, A case, $T_{\rm{eff}}$~=~[7000, 10000, 18000, 30000] K,
log~$g$~=~[1.8, 3.8] cm/s$^{2}$, [Fe/H]~=~[-0.5, 0.5] dex
 \item  F, G K case, $T_{\rm{eff}}$~=~[4000, 5600, 7200] K,
log~$g$~=~[1.8, 3.8] cm/s$^{2}$, [Fe/H]~=~[-1.7, -0.3] dex
 \item  M case, $T_{\rm{eff}}$~=~[3100, 3600, 4100] K,
log~$g$~=~[1.0, 4.0] cm/s$^{2}$, [Fe/H]~=~[-0.5, 0.0] dex
\end{itemize}

The adopted solution, i.e. absolute minimum, is the best of those
obtained with different guesses.

Because the S/N in the ELODIE spectra drops notably in the blue,
especially for cool stars, we restricted the fit to the wavelength
range 4400-6800~\AA{} (though using the whole 3900-6800~\AA{} gives
consistent results). Among the 1273 CFLIB stars, 885 have a complete
coverage of the wavelength range (neglecting small gaps of less than
50 ~\AA). The other stars contain gaps in the coverage. The bad
pixels (e.g. emission lines, cosmic ray, bad sky subtraction,
telluric lines or bad ones due to the instrument errors etc.) are
automatically rejected during the fitting process by \ulyss{}, using
the clipping algorithm applied iteratively on the residuals to the
fit ({\sc /CLEAN} option).

\onltab{2}{
\longtab{2}{
\setlength{\tabcolsep}{4.5pt}
\begin{longtable}{l|l|rr|rr|rr|c}
\caption{\label{meas.tbl}Adopted atmospheric parameters for the 1273 CFLIB stars}\\
\hline\hline
Name\tablefootmark{a} &
       Sp.Type\tablefootmark{b} &
                 $T_{\rm eff}$ & error & $log(g)$ & error & [Fe/H] & error & Flag \\
     &         & \multicolumn{2}{c|}{(K)}            & \multicolumn{2}{c|}{($cm\,s^{-2}$)} & \multicolumn{2}{c|}{(dex)} & \\
\hline

\endfirsthead
\caption{continued.}\\
\hline\hline

Name\tablefootmark{a} &
       Sp.Type\tablefootmark{b} &
                 $T_{\rm eff}$ & error & $log(g)$ & error & [Fe/H] & error & Flag \\
     &         & \multicolumn{2}{c|}{(K)}            & \multicolumn{2}{c|}{($cm\,s^{-2}$)} & \multicolumn{2}{c|}{(dex)} & \\
\hline
\endhead
\hline
\endfoot
HD000249      &K1IV          & 4757& $\pm$  28  &2.88 & $\pm$ 0.09  &-0.29 & $\pm$ 0.06 & 0  \\
HD000358      &B8IVmnp...    &12307& $\pm$ 180  &4.14 & $\pm$ 0.07  & 0.38 & $\pm$ 0.09 & 0  \\
HD000400      &F8IV          & 6215& $\pm$  45  &4.18 & $\pm$ 0.07  &-0.24 & $\pm$ 0.07 & 0  \\
HD000417      &K0III         & 4858& $\pm$  33  &2.45 & $\pm$ 0.10  &-0.30 & $\pm$ 0.08 & 0  \\
HD000693      &F5V           & 6221& $\pm$  41  &4.14 & $\pm$ 0.06  &-0.34 & $\pm$ 0.06 & 0  \\
HD000886      &B2IV          &21397& $\pm$ 382  &3.84 & $\pm$ 0.04  &-0.06 & $\pm$ 0.03 & 0  \\
HD001227      &G8II-III      & 5068& $\pm$  41  &2.81 & $\pm$ 0.10  & 0.00 & $\pm$ 0.09 & 0  \\
HD001406      &K3III         & 4349& $\pm$  20  &2.61 & $\pm$ 0.09  &-0.13 & $\pm$ 0.06 & 0  \\
\hline
\end{longtable}
\tablefoot{The last column is a flag indicating the origin of the adopted parameters,
`0', standard fit of the CFLIB spectrum;
`1', fit of the CFLIB spectrum starting from 3900~\AA;
`2', from literature;
`3', \ulyss{} fit of ELODIE spectra;
`4', improved fit, see the text.
The internal errors are given only when Flag\,=\,0 or 1.
\\
\tablefoottext{a}{Designation of the stars in the original CFLIB list \citep{cflib}}.
\tablefoottext{b}{Stellar spectral classification inherited from \citet{cflib},
and updated with recent classifications for the literature.}
} 
} 
} 

\section{Determination of the atmospheric parameters}

In this section, we present the determination of the atmospheric
parameters and the comparison with previous publications, separately
for the three groups of stars with spectral types FGK, OBA, and M.

Table~\ref{meas.tbl} lists the atmospheric parameters of the 1271
CFLIB stars determined with \ulyss{}.

We could not determine the atmospheric parameters of HD\,156164
(A3IV) because its observed wavelength range (6775-8648~\AA) does
not overlap the range of ELODIE. We rejected also HD\,25329
(K1V...), whose CFLIB spectrum is apparently corrupted. We suspect
that the range around $H_\beta$ corresponds to another star. It may
be a pointing error, possibly because of the high proper motion of
this star; the A0 star HD\,279382, located 3 arcmin away, may have
been observed for the $H_\beta$ setup.

\addtocounter{table}{1}  

\subsection{F, G, and K stars}
\label{sec.fgk}

The majority of the stars in both CFLIB and ELODIE libraries have
spectral types F, G and K; 958 of the 1273 CFLIB stars belong to
these types.  On the basis of our analysis, we also assigned to this
group five stars without a classification. These are BD+09~3063,
BD+09~3223, BD+18~2890, G~102-20 (=~BD+12~853) \& G~46-31
(=~HIP~45554).

There are five stars with poor quality spectra or with missing regions:
HD\,186408 (G1.5Vb) has
very low S/N in the range 4750-6070~\AA{} and we fit only the range
6070-6800~\AA. For HD\,140283 (F3) we fit the range 4780-6800~\AA.
For HD\,18474 (G5:III...) we fit the range 6010-6800~\AA.
HD\,224458 (G8III) has a large gap 4780.6-6812.2~\AA{} and the fit
is limited to the range 3900-4780.6~\AA. Finally, BD+09~3063's coverage is
similarly limited to 3900-4782.6~\AA.

\begin{figure*}
\centering
\includegraphics[width=12cm]{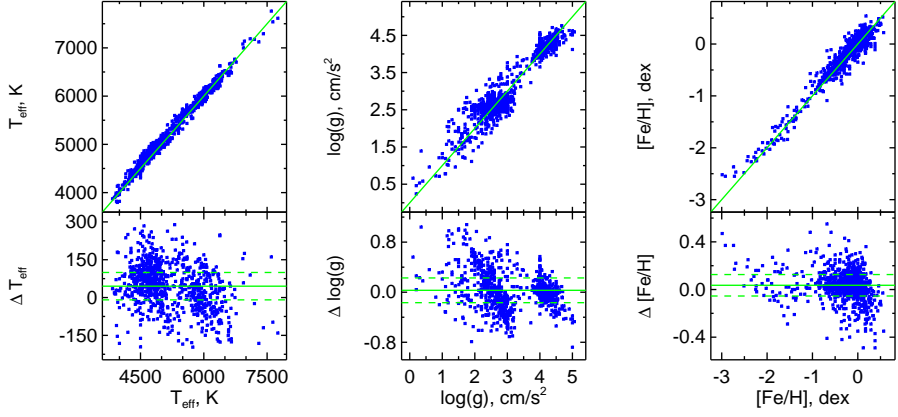}
\caption {Comparison of the atmospheric parameters determined by us
(ordinates of the upper panels) with those from Valdes (abscissas).
818 FGK stars are plotted and 46 temperature, eight log~$g$ and five
[Fe/H] outliers are not displayed. In the lower panels, we plot the
differences between the values from \ulyss{} and the literature
(here Valdes). The green lines in the upper panels are the 1 to 1
ratios. In the lower panels, the continuous green lines mark the
mean bias, and the dashed lines are the 1-$\sigma$ deviation
(computed after the rejection of the $T_{\rm{eff}}$ outliers).  The
same convention is used from Fig.~\ref{fig.comp12} to
\ref{fig.compm4} for other literature sources. }
\label{fig.compcflib}
\end{figure*}

Figure~\ref{fig.compcflib} compares our determinations of the
atmospheric parameters with those compiled by \citet[][hereafter,
Valdes]{cflib} in the original CFLIB release, for the 877 FGK stars
for which Valdes gives the three parameters. The statistics shown in
Table~\ref{tb.fgk} emphasize the heterogeneity of the Valdes
compilation. We excluded 46 $T_{\rm{eff}}$, 8 log~$g$ and 5 [Fe/H]
outliers.

In Appendix~\ref{appendix:comparisonfgk} we compare our
determinations to the ELODIE 3.2 internal estimations and to 15
other previous studies with a significant number of F, G, and K
stars in common with CFLIB. We determined the mean biases and the
rms dispersion between these datasets and our series of measurement
for the three parameters, after rejecting the objects whose
$T_{\rm{eff}}$ deviate by more than 3-$\sigma$. For the comparison
with ELODIE 3.2 and \citet{nord04} we also excluded some log~$g$ and
[Fe/H] outliers.

The summary of these comparisons is reported in Table~\ref{tb.fgk}.
The most significant outliers are listed in Table~\ref{tb.outlier}
and are discussed in the appendix.

\begin{table*}
\caption{\label{tb.fgk} External comparison of the atmospheric
parameters of F, G, and K stars.}
\begin{tabular}{l|cc|cc|cc|c|c}
\hline\hline
Reference & \multicolumn{2}{c|}{$\Delta$$T_{\rm{eff}}$ (K)} & \multicolumn{2}{c|}{$\Delta$log$~g$ ($cm\,s^{-2}$)} & \multicolumn{2}{c|}{$\Delta$[Fe/H] (dex)} & No. of Stars\tablefootmark{a}  & Sp.Type \\
          & mean & rms & mean & rms & mean & rms & /clipped & \\
 \hline                                                                                 
Valdes et al. (2004)     &   45.1 &  98.7 &  0.03 & 0.30 &  0.04 & 0.13 & 877/46/8/5 & F,G,K \\
ELODIE 3.2 I (2009)      &   17.4 &  41.4 &  0.04 & 0.10 &  0.03 & 0.07 & 387/7/1/1  & F,G,K  \\ 
Valenti \& Fischer (2005)&  -27.4 &  78.5 & -0.11 & 0.16 & -0.04 & 0.06 & 106/0  & F,G,K  \\ 
Soubiran et al. (2008)   &   39.9 &  75.1 &  0.09 & 0.27 &  0.03 & 0.10 & 175/8  & F,G,K  \\ 
Edvardsson et al. (1993) &  -14.8 &  52.2 & -0.02 & 0.12 &  0.04 & 0.07 & 129/10 & F,G    \\ 
Santos et al. (2004)     &  -46.3 &  74.4 & -0.06 & 0.15 & -0.02 & 0.07 & 23/2   & F,G,K  \\ 
Fuhrmann et al. (1998)   &  -27.9 &  41.5 & -0.04 & 0.10 & -0.01 & 0.03 & 7/0    & F,G    \\ 
da Silva et al. (2006)   &    5.0 &  32.4 &  0.11 & 0.17 &  0.01 & 0.06 & 17/1   & G,K    \\ 
Gray et al. (2001)       &  -52.9 &  91.8 &  0.06 & 0.28 &  0.03 & 0.11 & 46/4   & F,G    \\ 
Robinson et al. (2007)   &   -4.6 &  57.2 & -0.05 & 0.27 &  0.00 & 0.11 & 26/2   & F,G,K  \\ 
Nordstr\"{o}m et al. (2004)& 36.8 &  62.0 &  -    & -    &  0.03 & 0.08 & 269/15/0/2 & F,G,K  \\ 
Mishenina et al. (2006)  &   41.3 &  55.8 &  0.19 & 0.28 &  0.01 & 0.09 & 48/3   & G,K    \\ 
Kovtyukh et al. (2006)   &   42.5 &  64.4 &  -    & -    &  -    & -    & 69/5   & G,K    \\ 
Takeda (2007)            &  -24.3 &  71.3 &  0.00 & 0.13 & -0.02 & 0.06 & 117/9  & F,G,K  \\ 
Hekker (2007)            &  -10.8 &  62.2 & -0.23 & 0.29 &  0.01 & 0.07 & 132/5  & G,K    \\ 
Luck (2007)              &  -65.5 &  70.5 & -0.18 & 0.24 & -0.06 & 0.07 & 113/5  & G,K    \\ 
Sousa (2008)             &  -43.7 &  75.1 & -0.07 & 0.16 &  0.00 & 0.05 & 17/2   & F,G,K  \\ 
\hline
\end{tabular}
\tablefoot{
The mean $\Delta$ values are computed as our determination minus those of the literature.\\
\tablefoottext{a}{The first number is the total of spectra in
common, the second is the number of the rejected $T_{\rm{eff}}$
outliers. For the comparisons with ELODIE 3.2, Valdes, and
Nordstr\"{o}m, the third and fourth are the numbers of rejected
log~$g$ and [Fe/H] outliers .}}
\end{table*}

\subsection{O, B, and A Stars}

There are 260 O, B, and A stars in CFLIB. The determination of the
atmospheric parameters for hot stars is more delicate than for the
FGK stars for a couple of reasons. First, there are fewer spectral
features. Second, the spectra are often more complex because of
emission lines, of chemical peculiarities or of rotation. Third,
there are few reference measurements in the literature  and the
number of stars in the ELODIE library is even smaller. For these
reasons, the interpolator is less secure, and may be biased, in the
sense that the interpolated spectra may differ from the templates by
some systematics.

To study these systematics, we compare in
Appendix~\ref{appendix:comparisonoba} the absolute and internal
parameters of the ELODIE library. We also compare our determinations
with three other studies. The results are given in
Table~\ref{tb.oba}.

For two stars we excluded a significant region of the spectrum:
For HD\,206267 (O6e), the region 4785~-~5933~\AA{} was not observed, and for
HD\,33111 (A3III), the region 4679~-~6012~\AA{} has a low S/N.

\begin{table*}
\caption{Comparison of common O, B, and A type stars with other
references.}
\begin{tabular}{l|cc|cc|cc|c|c}
\hline\hline
Reference & \multicolumn{2}{c|}{$\Delta$$T_{\rm{eff}}$(\%)} & \multicolumn{2}{c|}{$\Delta$log$~g$ ($cm\,s^{-2}$)} & \multicolumn{2}{c|}{$\Delta$[Fe/H] (dex)} &  No. of Stars & Sp.Type \\
          &  mean & rms & mean & rms & mean & rms & /clipped & \\
 \hline
ELODIE 3.2 A vs. I (2009)& 1.33 & 6.32 & 0.07 & 0.25 & -0.01 & 0.21 & 293/19     & O,B,A \\
ELODIE 3.2 I (2009)      &-0.84 & 3.36 &-0.05 & 0.16 &  0.01 & 0.14 &  47/6      & O,B,A \\
Valdes et al. (2004)     & 1.53 &10.02 & 0.08 & 0.47 &  0.15 & 0.55 &  86/2      & O,B,A \\
                         & 1.53 &10.02 & 0.01 & 0.35 &  0.02 & 0.29 &  86/2/5/18 & O,B,A \\
Cenarro et al. (2007)    & 4.33 &12.14 & 0.17 & 0.58 &  0.23 & 0.60 &  38/1      & B,A   \\
                         & 4.33 &12.14 & 0.03 & 0.38 & -0.01 & 0.33 &  38/1/3/9  & B,A   \\
\hline
\end{tabular}
\label{tb.oba} \tablefoot{ For the comparison with Valdes et al.
(2004) and Cenarro et al. (2007), the statistics listed in the first
line are given after excluding the temperature outliers, and the
second line lists the statistics after clipping those red crosses in
log~$g$ and [Fe/H] panels respectively. Numbers displayed in the
column of `No. of Stars /clipped' show the clipped number in the
order of $T_{\rm{eff}}$, log~$g$ and [Fe/H]. }
\end{table*}

\subsection{M Stars}

\begin{table*}
\caption{Comparison of common M type stars with other references.}
\begin{tabular}{l|cc|cc|cc|c}
\hline\hline
Reference  & \multicolumn{2}{c|}{$\Delta$$T_{\rm{eff}}$ (K)} & \multicolumn{2}{c|}{$\Delta$log$~g$ ($cm\,s^{-2}$)} & \multicolumn{2}{c|}{$\Delta$[Fe/H] (dex)} & Star num/clip \\
           &  mean & rms & mean & rms & mean & rms & \\
 \hline
ELODIE 3.2 I (2009)   & -32.9 &  43.7 &  0.16 & 0.34 &  0.03 & 0.08 & 7    \\
Valdes et al. (2004)  & -60.9 & 122.9 & -0.08 & 0.32 & -0.12 & 0.44 & 18/2 \\
Cenarro et al. (2007) & -61.7 & 127.3 &  -    & -    &  -    & -    & 10   \\
Cayrel et al. (2001)  & -97.4 & 133.4 & -0.11 & 0.30 & -0.14 & 0.49 & 18   \\
\hline
\end{tabular}
\label{tb.m}
\end{table*}

There are 53 M, S, or C stars in CFLIB. For nine of them, the fit
with \ulyss{} was not successful, in the sense that the metalicity
reached the lower bound, [Fe/H]\,=\,-1.0~dex, set as a validity
limit of the models (for the M stars). After a critical review of
the literature and of the quality of our fit, we adopted the
parameters listed in  Table~\ref{tb.9m}.

\begin{table}
\caption{Nine late type stars,which \ulyss{} could not successfully
fit.}
\begin{tabular}{l|l|rrr|c}
\hline\hline
Name & Sp.Type & $T_{\rm eff}$ & $log(g)$   &  [Fe/H] & Ref. \\
     &         & (K)        & $cm\,s^{-2}$ &  (dex)  & \\
\hline
HD078712 & M6IIIase  & 3210 & 0.00 & -0.01 & A    \\ 
         &           & 3101 & 0.23 & -1.00 & 0    \\
         &           & 3210 & 0.00 & -0.11 & 1    \\
         &           & 3210 & 0.00 &  0.09 & 1    \\
HD084748 & M8IIIe    & 3070 & 0.78 & -1.00 & 0\&A \\ 
HD114961 & M7III     & 3014 & 0.00 & -0.81 & 2\&A \\ 
         &           & 3080 & 0.36 & -1.00 & 0    \\
HD126327 & M7.5      & 3000 & 0.00 & -0.58 & 2\&A \\ 
         &           & 3088 & 0.30 & -1.00 & 0    \\
HD148783 & M6III     & 3250 & 0.20 & -0.04 & A    \\ 
         &           & 3146 & 0.45 & -1.00 & 0    \\
         &           & 3250 & 0.20 & -0.01 & 3    \\
         &           & 3250 & 0.20 & -0.06 & 4    \\
HD177940 & M7IIIevar & 3092 & 0.59 & -1.00 & 0\&A \\ 
HD187796 & S...      & 3148 & 0.49 & -1.00 & 0\&A \\ 
HD196610 & M6III     & 3100 & 0.39 & -1.00 & 0\&A \\ 
HD197812 & M5Iab:    & 3389 & 0.32 & -1.00 & A    \\ 
         &           & 3119 & 0.32 & -1.00 & 0    \\
         &           & 3389 &      &       & 5    \\
\hline
\end{tabular}
\tablebib{
 (A) Adopted solution;
 (0) \ulyss{} fit to the CFLIB spectrum;
 (1) \citet{smith86}; (2) \citet{Jones};
 (3) \citet{rami00}; (4) \citet{carr00}; (5) \citet{dyck98}.
} \tablefoot{ First column is the star name. Second column is the
stellar spectral type. Third to fifth column are the three
atmospheric parameters. Sixth lists the source of the measurements.
For HD\,78712 we adopt the averaged parameters from \cite{smith86}'s
two solutions, for HD\,114961 and 126327 we adopt values from
\cite{Jones}, for HD\,148783 we adopt the averaged values of
\cite{rami00} and \cite{carr00}, for HD\,197812  we use the
$T_{\rm{eff}}$ value from \citet{dyck98}, and adopt log~$g$ and
[Fe/H] value of \ulyss{} determination.  For the remaining four
stars, as no other reference could be found, we adopt our
determinations in the final table. } \label{tb.9m}
\end{table}

In Appendix~\ref{appendix:comparisonm} we compare our measurements
with other determinations from four other compilations. The
corresponding statistics is summarized in Table~\ref{tb.m}.

\section{Discussion and conclusion}

Nearly 300 of the 1273 stars in the Indo-US stellar spectral library
(CFLIB) had one or more of the atmospheric parameters,
$T_{\rm{eff}}$, log~$g$, or [Fe/H], unknown.

\begin{figure}
\includegraphics[width=9cm]{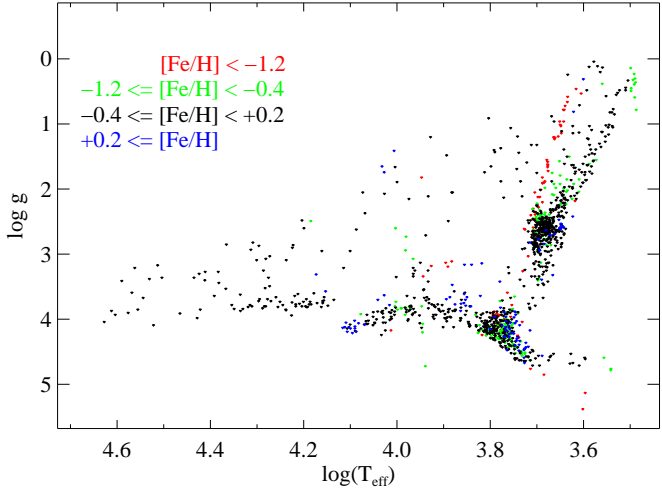} 
\caption {Distribution in the log($T_{\rm{eff}}$) - log~$g$ plane of
the adopted atmospheric parameters for the 1271 CFLIB stars. The
color of the symbols distinguishes the different metalicity classes.
} \label{fig.distri_tg}
\end{figure}

\begin{figure}
\includegraphics[width=9cm]{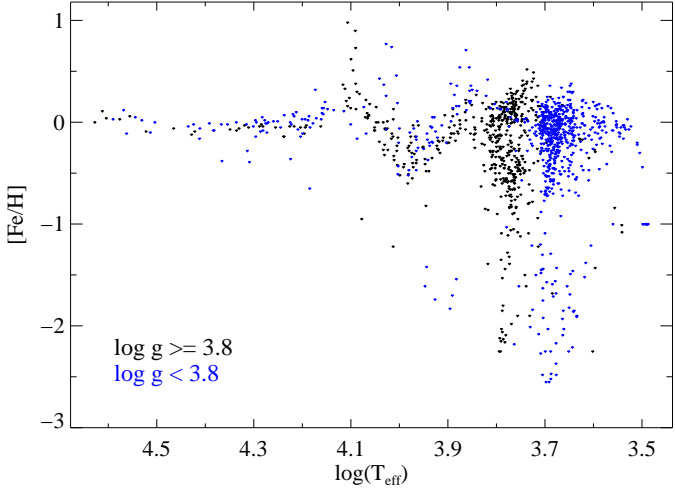} 
\caption {Distribution in the log($T_{\rm{eff}}$) - [Fe/H] plane
of the adopted atmospheric parameters for the 1271 CFLIB stars. The dwarfs
are in black and the giants in blue.
}
\label{fig.distri_tm}
\end{figure}

We have determined the atmospheric parameters for 1271 (out of 1273)
CFLIB stars using the \ulyss{} package and the interpolator of the
ELODIE library. ULySS  optimizes the usage of all the signal
information inside the spectrum, allows taking into account of the
errors on each bin of the spectrum and is insensitive to the shape
of the continuum. This method determines all free parameters within
a single minimization, in order to properly handle the degeneracy
between some parameters, e.g., temperature and metalicity. The
instrumental and physical broadening of the spectra is matched with
a Gaussian convolution. The distribution of the adopted parameters
is presented on Figs.~\ref{fig.distri_tg}~\&~\ref{fig.distri_tm}.

Based on comparisons with several previous studies we conclude that
our method is robust. We derived the intrinsic external accuracy.
For the 958 F, G, and K stars, the precisions on $T_{\rm{eff}}$,
log~$g$, and [Fe/H] are respectively 43~K, 0.13, and 0.05~dex. For
the 53 M stars they are 82~K, 0.22, and 0.28~dex and for the 260 O,
B, and A stars the relative precision on $T_{\rm{eff}}$ is 5.1\%,
and on log~$g$, and [Fe/H] the precisions are  0.19 and 0.16~dex
respectively.

The external comparisons also allow us to probe the existence of
biases. For the FGK stars the various comparisons summarized in
Table~\ref{tb.fgk} indicate biases in the ranges $-65$ to 44~K,
$-0.2$ to 0.1~dex and $-0.06$ to 0.04~dex, for temperature, gravity
and metallicity respectively. For OBA stars (Table~\ref{tb.oba}),
they are in the ranges $-1$ to 4~\%, $-0.05$ to 0.03~dex and $-0.01$
to 0.02~dex, and for M stars (Table~\ref{tb.oba}) $-100$ to $-30$~K,
$-0.1$ to 0.1~dex and $-0.13$ to 0.03~dex. These individual biases
are symmetrical around zero, and we have no indication for a
systematics in our measurements.

These determinations are particularly reliable and accurate for the
FGK stars. But they appeared also valuable for both the OBA and M
stars. The main limitations are for the low-metalicity blue
horizontal branch stars and the stars cooler than 3300 K. In the
first case, the program tends to over-estimate the metalicity. An A0
star with [Fe/H]~$\sim~-1.5$ would be retrieved at
[Fe/H]~$\sim~-0.8$. In the second case, the program often converges
toward a low-metalicity solution: A solar metalicity M star may be
found at the low-metalicity boundary of the model. Both of these
effects are certainly owing to the poor coverage of the ELODIE
library in these regions of the parameter space. Also, the physical
description of the hot and cool stars cannot be generally expressed
by the three atmospheric parameters plus a Gaussian broadening. More
elaborated models may improve the behavior of the program for these
special stars, or a non-parametric approach, like TGMET, may perform
better.

We will use this set of atmospheric parameters to re-calibrate the
fluxes in the CFLIB library.  Then, after homogenizing and
correcting the line-spread functions, we will prepare high-quality
stellar population models.

\begin{acknowledgements}

We thank R. Peletier (the referee), C. Soubiran and Th. Morel for comments
which helped to improve the manuscript.
PP is grateful to Indo-French Astronomy Network (IFAN) for a visit
to India where this work was initiated, and for the support from the
French {\it Programme National Cosmologie et Galaxies} (PNCG).
YW acknowledges a grant from the China Scholarship Council (CSC) under No. 2007104275 and a
funding from Natural Science Foundation of China (NSFC) under No.10973021.
She also thanks Prof. Y.H. Zhao and Prof. A.L. Luo for providing 
support from the Chinese Academy of Sciences (LAMOST).
RG, HPS and MK thank CRAL, Observatoire de Lyon, Universit\'{e} Claude
Bernard for Invited Professorships.

\end{acknowledgements}

\bibliographystyle{aa} 
\bibliography{tgmbib}  

\begin{appendix}

\section[]{External comparisons for FGK stars} \label{appendix:comparisonfgk}

In this appendix we present the detailed comparisons between our measurements
and 16 previous studies. The main outliers are individually discussed.


\subsection{Comparison with ELODIE 3.2}
\label{sec.comp12}

\begin{figure*}
\centering
\includegraphics[width=12cm]{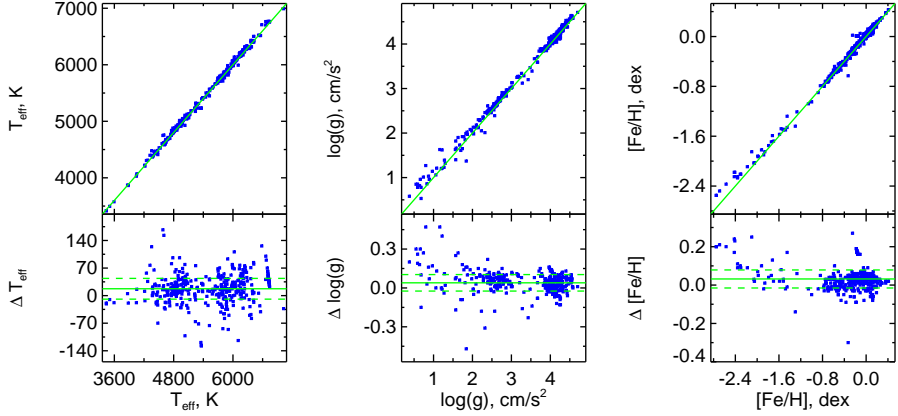}
\caption {Comparison of the atmospheric parameters determined by us
(ordinates, upper panels) with those from ELODIE 3.2 (abscissas). We
plot 378 F, G, and K observations in common between the two data
sets, seven outliers are not displayed. The axes are as in
Fig.~\ref{fig.compcflib}. }
 \label{fig.comp12}
\end{figure*}

There are 206 stars in common between CFLIB and the ELODIE library,
corresponding to a total of 387 ELODIE spectra. The comparison with
the internal parameters of ELODIE is shown in Fig.~\ref{fig.comp12}.
The ELODIE parameters were determined with the interpolator computed
on the continuum normalized spectra, while for the present study we
used the flux-calibrated interpolator.

The dispersion between the two series of measurements of
$T_{\rm{eff}}$ is 42~K (see Table~\ref{tb.fgk}). Assuming that the
uncertainty is evenly shared between the two series, this implies an
intrinsic precision of  $\sim$30~K, in agreement with the results of
Sect.~\ref{sec.consistency}. This precision is well within the
errors quoted in the literature (50 to 70~K) and we conclude that we
can rely on our method. The precisions on log~$g$ and [Fe/H] are
respectively 0.08 and 0.06 dex.

We detected seven  $T_{\rm{eff}}$, one log~$g$ and two [Fe/H]
outliers that we excluded from the statistics. For HD\,22468
(K2:Vnk), we obtained a temperature $\sim$~390~K cooler than the one
in ELODIE, $T_{\rm{eff}} \sim$~5136~K, (average of two spectra). Our
fit of the two ELODIE spectra returns a mean
$T_{\rm{eff}}$~=~4771~K, consistent with our estimation. This star
is an active RS~CVn variable with strong emission lines (CAII K \&
H, Balmer lines) and a broad asymmetric cross-correlation peak
reflecting the duplicity. The main component is a K1/2 star
\citep{gray06}, for which a temperature around 4800~K is reasonable.
We keep our determination in the final table.

For the second outlier, HD\,72946, with three ELODIE observations,
we find a temperature $\sim$ 360~K hotter than the ELODIE internal
determinations, $T_{\rm{eff}}$~=~5655~K, or the \ulyss{} fit of the
ELODIE spectra, $T_{\rm{eff}}$~=~5664~K. As the CFLIB observation
misses the red part of the fitting wavelength range, we remade the
fit starting from 3900~\AA, and obtained $T_{\rm{eff}}$~=~5624~K. We
adopt this latter solution. For the third and fourth outliers
HD\,183085 and 45674 we determined temperatures $\sim$~260~K warmer
than the ELODIE internal estimation. However, as our solution is
consistent with the \ulyss{} fit of the ELODIE spectra, we adopt it.

There is also an obvious log~$g$ and [Fe/H] outlier, HD\,178266
(K5). We find log~$g$~=~2.68~dex and [Fe/H]~=~0.10~dex, while ELODIE
3.2 quotes log~$g$~=~1.76~dex and -0.50~dex (absolute) and 1.81~dex
and -0.49~dex (internal). In ELODIE version 1 \citep{PS01}, the star
was on the dwarf sequence: log~$g$~=~5.27~dex. Fitting the ELODIE
spectrum with \ulyss{}, we obtain log~$g$~=~2.69~dex and
[Fe/H]~=~0.10~dex, in agreement with our measurement of the CFLIB
spectrum, which we finally adopt.

Finally, there is another metalicity outlier,  HD\,22211 (G0). We
obtained [Fe/H]~=~0.20~dex while CFLIB quotes -0.31~dex (from ELODIE
version 1) and ELODIE 3.2 gives -0.29~dex. Fitting the ELODIE
spectrum, we get [Fe/H]~=~0.14~dex, close to our determination,
which we therefore adopt.

\subsection{Comparison with Valenti \& Fisher (2005)}

\cite{VF05} published a catalog of stellar properties for 1040
nearby F, G, and K dwarfs (1944 spectra) observed at a resolution
R\,$\sim$\,70\,000 by various planet search programs. Directly
fitting  the observations rather than equivalent widths with
synthetic spectra yielded the effective temperature, surface
gravity, metalicity, and projected rotational velocity. They adopted
a \cite{kurucz92} grid of model atmospheres generated by the ATLAS9
program \citep{kurucz93a}. Their precisions are 44~K for
$T_{\rm{eff}}$, 0.06~dex for log~$g$ and 0.03~dex for [Fe/H].
Beside, they compared their measurements with \cite{ed93},
\cite{fuhr97}, \cite{fuhr98}, \cite{santos04} and \cite{prieto04},
and found mean external errors of 72~K in $T_{\rm{eff}}$, 0.13~dex
in log~$g$, and 0.06~dex in [Fe/H].

\begin{figure*}
\centering
\includegraphics[width=12cm]{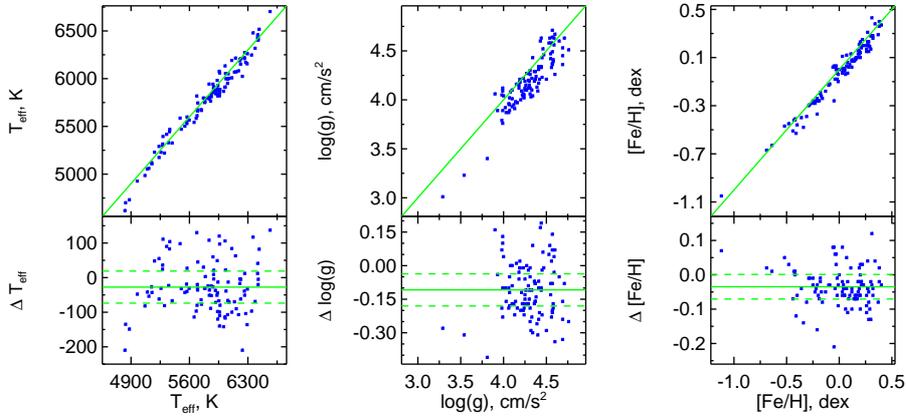}
\caption {Comparison of the atmospheric parameters determined by
us with those from \cite{VF05}. We plot 106 F, G, and K dwarfs
in common between the two data sets.
The axes are as in Fig.~\ref{fig.compcflib}.
} \label{fig.comp1}
\end{figure*}

We have identified 106 CFLIB stars in common with this work and show
the comparison in Fig.~\ref{fig.comp1}. The two series are
consistent (see Table \ref{tb.fgk}), without outliers exceeding
$3~\sigma$ in $T_{\rm{eff}}$. Our determinations are within the
error estimates given by \cite{VF05}.

\subsection{Comparison with Soubiran et al. (2008)}
\label{sec.comp2}

\citet{soub08} assembled two lists of mostly clump giants:
368 nearby stars (d$<$100~pc) and 523 distant stars
(d$<$1~kpc) in the direction of the North Galactic pole. The
atmospheric parameters of the former group were either estimated
from their spectroscopic observations at high resolution and high
signal-to-noise ratio or compiled from the literature, including in
particular \citet{mish06}. For the latter group, they were determined with
the TGMET program \citep{Katz98} using ELODIE spectra.

\begin{figure*}
\centering
\includegraphics[width=12cm]{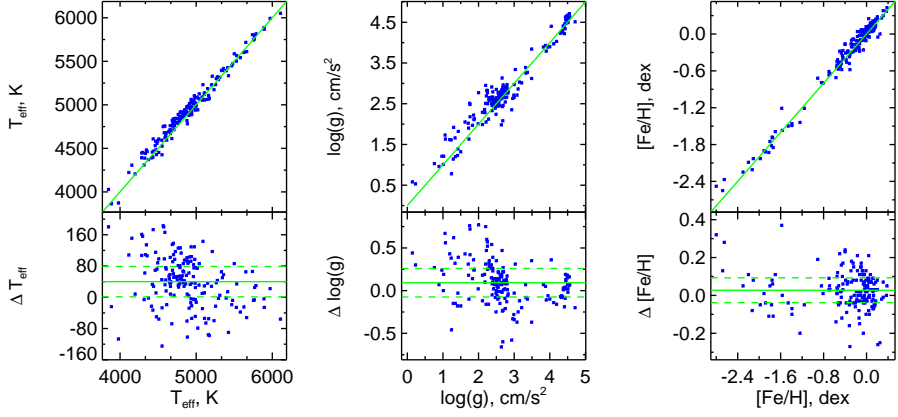}
\caption {Comparison of the atmospheric parameters determined by us
with those from \cite{soub08}. We plot 167 F, G, and K stars in
common between the two data sets, eight outliers are not displayed.
The axes are as in Fig.~\ref{fig.compcflib}.}
 \label{fig.comp2}
\end{figure*}

We identified 175 stars in common with this
reference, and we present the corresponding comparisons in
Fig.~\ref{fig.comp2} after excluding eight $T_{\rm{eff}}$ outliers.

For the first outlier HD\,48329 (G8Ib), we find a $\sim$\,360~K
warmer temperature. This star is one of the absolute calibrators in
\citet{soub08}, who adopted an average between the measurements by
\cite{luck82}, \cite{SL87} and \cite{mall98}. Our temperature,
4662~K, is consistent with \cite{luck82}'s, 4624~K, and we keep our
original measurements.

We found the second outlier, HD\,72946, 324~K warmer.
As the CFLIB spectrum misses the red range, we remade the fit
including the blue from 3900~\AA. This reduced the discrepancy to 80~K,
in agreement also with the ELODIE internal determination.
We adopted the results from our revised fit.

For the third outlier, HD\,197964, our $T_{\rm{eff}}\,=\,4762$~K is
$\sim$~230~K cooler. This star is in the TGMET reference library,
and the parameters given in \citet{soub08} are an average between
\cite{kyrol86} and \cite{mcwill90}. The latter is from a photometric
calibration, while the former, from a spectroscopic analysis, agrees
with our determination and is also consistent with the ELODIE
internal value. We adopt our original determination.

For the fourth (HD\,219134) and fifth (HD\,32147) outliers, our
fitted temperatures are $\sim$ 200~K cooler. For HD\,219134, the
\citet{soub08} temperature is an average between measurements by
\citet{str70}, 4710~K, \citet{oinas77}, 4710~K and \citet{robin07},
4963~K. Our $T_{\rm{eff}}\,=\,4700~K$ agrees with the former two and
is consistent with the ELODIE internal measurements (for two
spectra). For HD\,32147, the \citet{soub08} temperature is an
average between \cite{felt98} and \cite{thor00}, both from
photometric calibrations. Our determined temperature,
$T_{\rm{eff}}$\,=\,4617~K, is consistent with the ELODIE internal
value. We adopt our original determinations for these two stars.

The sixth outlier is HD\,40460 (K1III). Our $T_{\rm{eff}}$ is
263~K warmer. The \citet{soub08} temperature is a spectroscopic determination
from \cite{cott86}.

The two ELODIE internal determinations for this star are consistent with
our result and with the \ulyss{} fit of these spectra.
Therefore we adopt our original estimations.

For the seventh outlier HD\,16458 (G8p...), our $T_{\rm{eff}}$~=~4410~K
is $\sim$\,160~K cooler.
The parameters in \citet{soub08} are averages between
\citet{sneden81}, $T_{\rm{eff}}$\,=\,4500~K,
\citet{tomkin83}, 4600~K, \citet{smith84}, 4800~K, and
\cite{fern90}, 4500~K.
As the blue range of the spectrum has a good S/N, we re-fitted the
spectrum starting from 3900~\AA{} and obtained $T_{\rm{eff}}$~=~4484~K,
in agreement with both \cite{fern90} and the ELODIE internal
measurement. In Table~\ref{meas.tbl} we adopt our modified solution.

For the last outlier HD\,204867, our $T_{\rm{eff}}$~=~5705~K is
243~K warmer than \citet{soub08}, which is an average between
\cite{luck82}, 5362~K, and \cite{foy81}, 5478~K. The ELODIE internal
parameters are consistent with our results which we therefore adopt
in Table~\ref{meas.tbl}.

\subsection{Comparison with \cite{ed93}}

\citet{ed93} studied the atmosphere of 189 nearby field F and G
stars to provide observational constraints on the evolution of the
Galactic disk. The temperature and gravity were determined with
Str\"{o}mgren photometry, and the [Fe/H] metalicity and detailed
abundances were estimated to a precision of 0.05~dex with LTE
atmosphere models.

\begin{figure*}
\centering
\includegraphics[width=12cm]{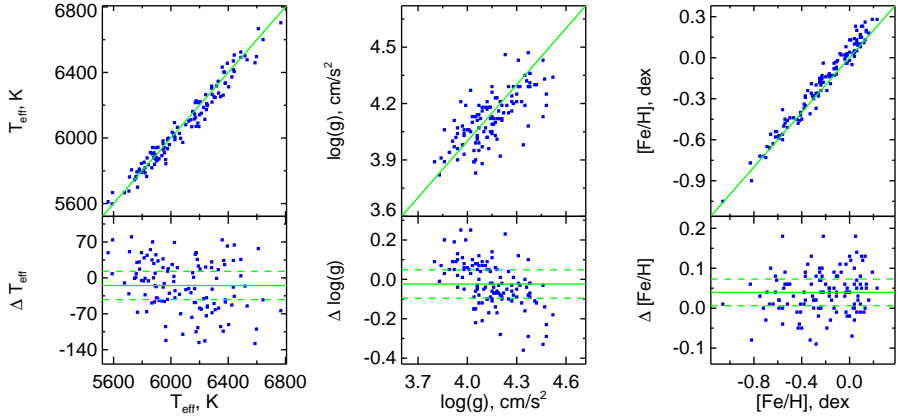}
\caption {Comparison of the atmospheric parameters determined by
us with those from \cite{ed93}. We plot 119 F and G stars in
common between the two data sets, 10 outliers are not displayed. The
axes are as in Fig.~\ref{fig.compcflib}.}
 \label{fig.comp3}
\end{figure*}

The CFLIB has 129 stars in common with this database and the results
of the comparison are shown in Fig.~\ref{fig.comp3}. Ten
$T_{\rm{eff}}$ outliers were excluded. As seen from Table
\ref{tb.fgk}, the deviations in the parameters are small between the
two studies.

For seven of the ten outliers we find a temperature around 170~K cooler
and, for the rest of them, about 100~K warmer. Six of these outliers
belong also to the ELODIE library, where their internal measurements
are consistent with the present determinations. The other four
$T_{\rm{eff}}$ outliers are discussed below.

For HD\,175317, our estimated temperature is $\sim$~170~K
cooler than the one given by \cite{ed93}, and there is also a
discrepancy of 0.20~dex in [Fe/H].
\cite{bala90} gives photometric determinations of the parameters
which are consistent with ours.
\cite{grat96} give  $T_{\rm{eff}}$\,=\,6517~K, only 38~K
warmer than us.
For HD\,102634, our estimated temperature is 177~K cooler
than \cite{ed93}'s, but our parameters agree with the measurements
of \citet{grat96}.
For HD\,220117 and 168151, our temperature is consistent with
\citet{boes90} and \citet{grat96}.
We adopt our original measurements for these four stars.

\subsection{Comparison with Santos et al. (2004)}

\citet{santos04} derived stellar metalicities and other parameters
from a detailed R~$\sim$~50\,000 spectroscopic analysis of a sample
of 139 stars known or suspected to be orbited by planetary mass
companions.

The errors were estimated to be on the order of 50~K in
$T_{\rm{eff}}$, 0.12~dex in log~$g$ and 0.05~dex in metalicity. The
obtained stellar parameters were found to be compatible, within the
errors, with the values derived by others. In particular, the
derived surface gravities were only on average 0.03~dex different
from trigonometric estimates based on Hipparcos parallaxes.

\begin{figure*}
\centering
\includegraphics[width=12cm]{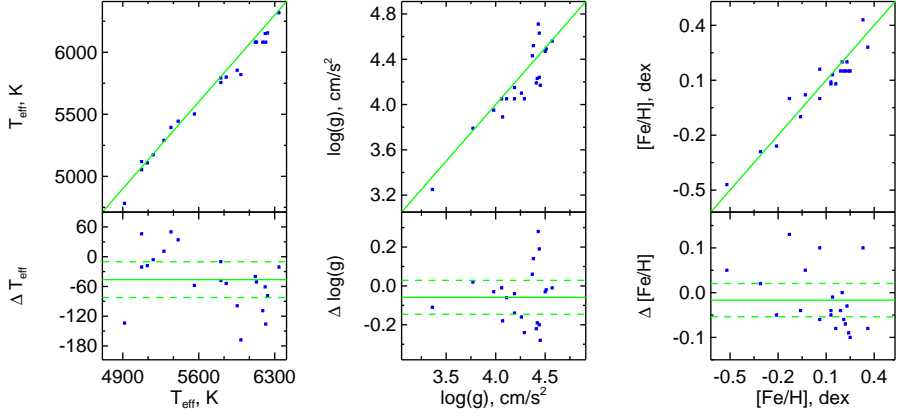}
\caption {Comparison of the atmospheric parameters determined by us
with those from \cite{santos04}. We plot 21 F, G, and K stars in
common between the two data sets, two outliers are not displayed.
The axes are as in Fig.~\ref{fig.compcflib}.}
 \label{fig.comp4}
\end{figure*}

We have a total of 23 F, G, and K stars common with this database.
Figure~\ref{fig.comp4} shows the comparison after excluding the two
outliers HD\,137759 and HD\,19994.

For HD\,137759, our $T_{\rm{eff}}$ is 250~K cooler, but our
atmospheric parameters are consistent with the ELODIE internal
determinations and are close to \cite{soub08}'s estimates,
especially the $T_{\rm{eff}}$ and [Fe/H]. For HD\,19994, our
$T_{\rm{eff}}$ is 209~K cooler, but closer to the photometric
estimate from \citet{ed93}. The internal measurements of the two
spectra from the ELODIE library are consistent with our results. For
both stars we adopt our measurements.

\subsection{Comparison with Fuhrmann et al. (1998)}

\begin{figure*}
\centering
\includegraphics[width=12cm]{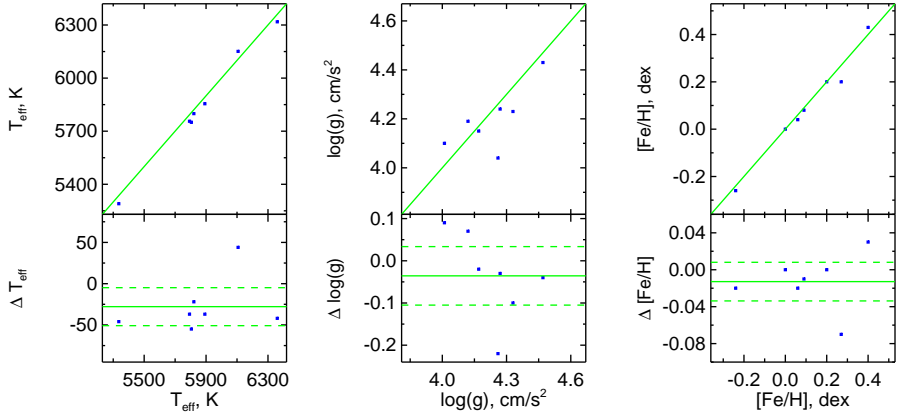}
\caption {Comparison of the atmospheric parameters determined by us
with those from \cite{fuhr98}. We plot seven F and G stars in common
between the two data sets. The axes are as in
Fig.~\ref{fig.compcflib}.}
 \label{fig.comp5}
\end{figure*}

\cite{fuhr98} used a $\rm H_{\alpha}$ and $\rm H_{\beta}$
line-fitting procedure to derive the effective temperatures of F and
G stars. Figure~\ref{fig.comp5} presents the comparison for the
seven stars in common with this sample. The agreement is good.

\subsection{Comparison with \citet{silva06}}

\Citet{silva06} give a detailed spectroscopic analysis of 72
evolved stars observed at the resolution R~=~50\,000. The
atmospheric parameters were obtained by adjusting the measured
equivalent widths to LTE atmosphere models, imposing excitation and
ionization equilibrium. A direct comparison of their results with
those of other authors \citep{cayrel01} showed that their metalicity
is on average systematically higher by 0.07 dex, their temperature
by about 40~K, and gravity by about 0.13~dex. Their estimate of the
uncertainties are 70~K on $T_{\rm{eff}}$ , 0.2~dex on log~$g$
0.1~dex on [Fe/H].

\begin{figure*}
\centering
\includegraphics[width=12cm]{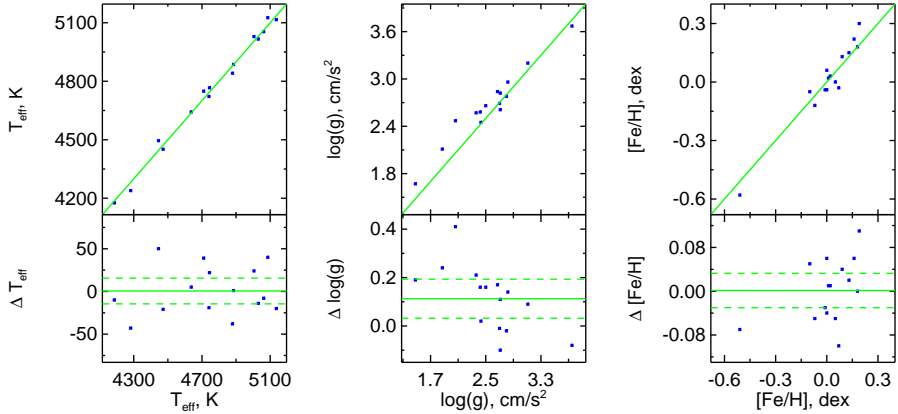}
\caption {Comparison of the atmospheric parameters determined by us
with those from \cite{silva06}. We plot 15 G and K stars in common
between the two data sets, two outliers are not displayed. The axes
are as in Fig.~\ref{fig.compcflib}.}
 \label{fig.comp6}
\end{figure*}

We found 17 G and K stars in common and show the comparison in
Fig.~\ref{fig.comp6}. We do not see the biases found by
\cite{silva06}. For HD\,99167, our estimated $T_{\rm{eff}}$ is 191~K
cooler, while the other two parameters also deviate significantly.
Our estimated $T_{\rm{eff}}$ and log~$g$ are consistent with
\citet{mcwill90} (discussed above in Sect.~\ref{sec.comp12}), but
our [Fe/H] deviates by 0.30~dex compared to both \cite{silva06} and
\cite{mcwill90}. The quality of our fit is satisfactory and we adopt
our determinations.

\subsection{Comparison with \cite{gray01}}

\citet{gray01} studied 372 late A, F, and early G type
stars with the aim of understanding the nature of the MK luminosity
classification for this range of spectral types.
They simultaneously fitted the optical spectra (1.8 \AA{} resolution)
and Str\"{o}mgren $uvby$ photometry against Kurucz ATLAS9 models.
They estimated the random external errors to 80~K on $T_{\rm{eff}}$,
0.1~dex on log~$g$ and 0.10 to 0.15 dex on [Fe/H].

\begin{figure*}
\centering
\includegraphics[width=12cm]{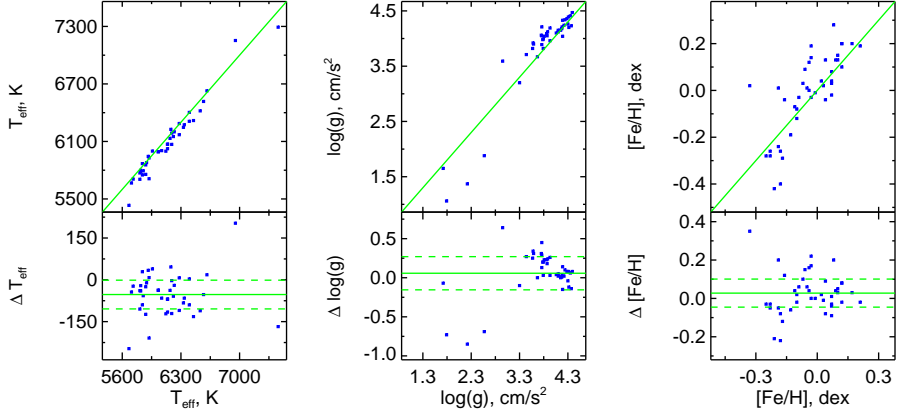}
\caption {Comparison of the atmospheric parameters determined by us
with those from \cite{gray01}. We plot 42 F and G stars in common
between the two data sets, four outliers are not displayed. The axes
are as in Fig.~\ref{fig.compcflib}.} \label{fig.comp7}
\end{figure*}

We present the comparison of the 46 stars in common with our sample
in Fig.~\ref{fig.comp7}, after excluding four $T_{\rm{eff}}$ outliers.

The star HD\,25291 (F0II) is a prominent outlier. The
$T_{\rm{eff}}$, log~$g$ and [Fe/H] from \cite{gray01} are 7050~K,
1.85~dex and -0.21~dex respectively. Our corresponding values are
7761~K, 2.49~dex and 0.14~dex, yielding large deviations of 711~K in
temperature, 0.64~dex in gravity and 0.35~dex in metalicity. This
star has been a subject of a number of investigations.
\cite{andri02} give $T_{\rm{eff}}$\,=\,6750~K, using photometric
data, and the log~$g$\,=\,1.00~dex was determined assuming the
FeI/II ionization equilibrium. \cite{giri05} gave spectroscopic
$T_{\rm{eff}}$\,=\,7250~K and log~$g$\,=\,1.50~dex. \cite{kov07}
gave 7497~K using line-depth ratios (the uncertainty is 5-30~K).
\cite{venn95} gave 7600~K and log~$g$\,=\,1.5~dex, from MgI/II
ionisation equilibrium and H$\gamma$ profile fit. All the above
references point to a mean log~$g~\sim$~1.50~dex, which seems
reliable. Fixing log~$g$ at this value and re-fitting the spectrum,
we find $T_{\rm{eff}}$\,=\,7390~K,  close to \cite{kov07}'s value
and more reasonable for a F0 star. In the final table we adopted
this latter solution.

The estimates for the other three $T_{\rm{eff}}$ outliers,
 HD\,194093, 36673 and 20902, can be
improved by fitting from 3900~\AA{} instead of 4400~\AA{}, because
the blue part of these spectra have a better quality. The internal
ELODIE measurements for HD\,194093 are consistent with the latter
estimations, and fitting the ELODIE spectrum with \ulyss{} also
gives a similar result. For HD\,20902, \cite{luck85} gave 6300~K,
cooler than \citet{gray01} by 260~K. In Table~\ref{meas.tbl} we
adopt our improved determinations.

\subsection{Comparison with \citet{robin07}}

\begin{figure*}
\centering
\includegraphics[width=12cm]{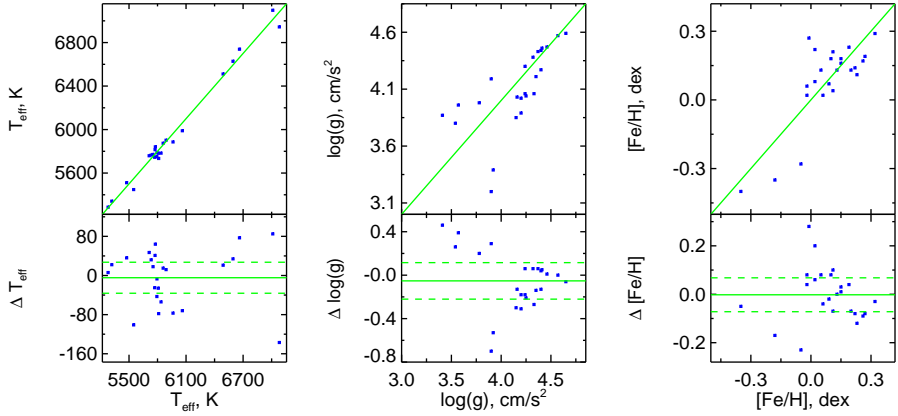}
\caption {Comparison of the atmospheric parameters determined by us
with those from \cite{robin07}. We plot 24 F, G, and K stars in
common between the two data sets, two outliers are not displayed.
The axes are as in Fig.~\ref{fig.compcflib}.}
 \label{fig.comp8}
\end{figure*}

\citet{robin07} reported atmospheric parameters for 1907 stars from
a low-resolution spectroscopic survey, designed to identify
metal-rich F, G, K dwarfs likely to harbor detectable planets.
[Fe/H], $T_{\rm{eff}}$ and log~$g$ were measured with the
calibrations of Lick indices presented in \citet{rob06}.
$T_{\rm{eff}}$ is given by a linear combination of Lick indices,
[Fe/H] by a linear combination of Lick indices and $T_{\rm{eff}}$
and log~$g$ by a linear combination of indices and $T_{\rm{eff}}$
plus one nonlinear term, $T_{\rm{eff}}$($\rm H_{\gamma F}$+$\rm
H_{\beta}$). The precision of these calibrations are cited to be
82~K, 0.13~dex and 0.07~dex on respectively  $T_{\rm{eff}}$, log~$g$
and [Fe/H]. We identified 26 common F, G, and K stars and show the
comparison in Fig.~\ref{fig.comp8} after clipping two $T_{\rm{eff}}$
outliers.

The first outlier, HD\,219134 (K3V), was discussed in
Sect.~\ref{sec.comp2}.  Our estimated temperature is 263~K cooler,
but is consistent with the ELODIE internal determination.
For the second, HD\,61295 (F6II), our $T_{\rm{eff}}$ is 196~K warmer,
but is consistent with the \cite{luck95a}'s photometric estimate.
We adopt our original determinations.

\subsection{Comparison with the Geneva-Copenhagen survey}

The Geneva-Copenhagen radial velocity survey of the solar
neighborhood \citep{nord04} has provided $T_{\rm{eff}}$ and [Fe/H]
for 16,682 nearby F and G dwarf stars from Str\"{o}mgren photometry.
The effective temperatures were determined from the
reddening-corrected $b-y$, $c_{\rm{1}}$, and $m_{\rm{1}}$ indices
and the calibration of \citet{alo96}. Their resulting temperatures
have a mean difference of only 3~K and a dispersion of 94~K compared
to \citet{bar02}. They compared their metalicities with the
homogeneous spectroscopic values from \citet{ed93} and
\citet{chen00}. They found  mean differences of 0.02 and 0.00~dex
and dispersions of 0.08 and 0.11~dex, respectively.

\begin{figure}
\includegraphics[width=9cm]{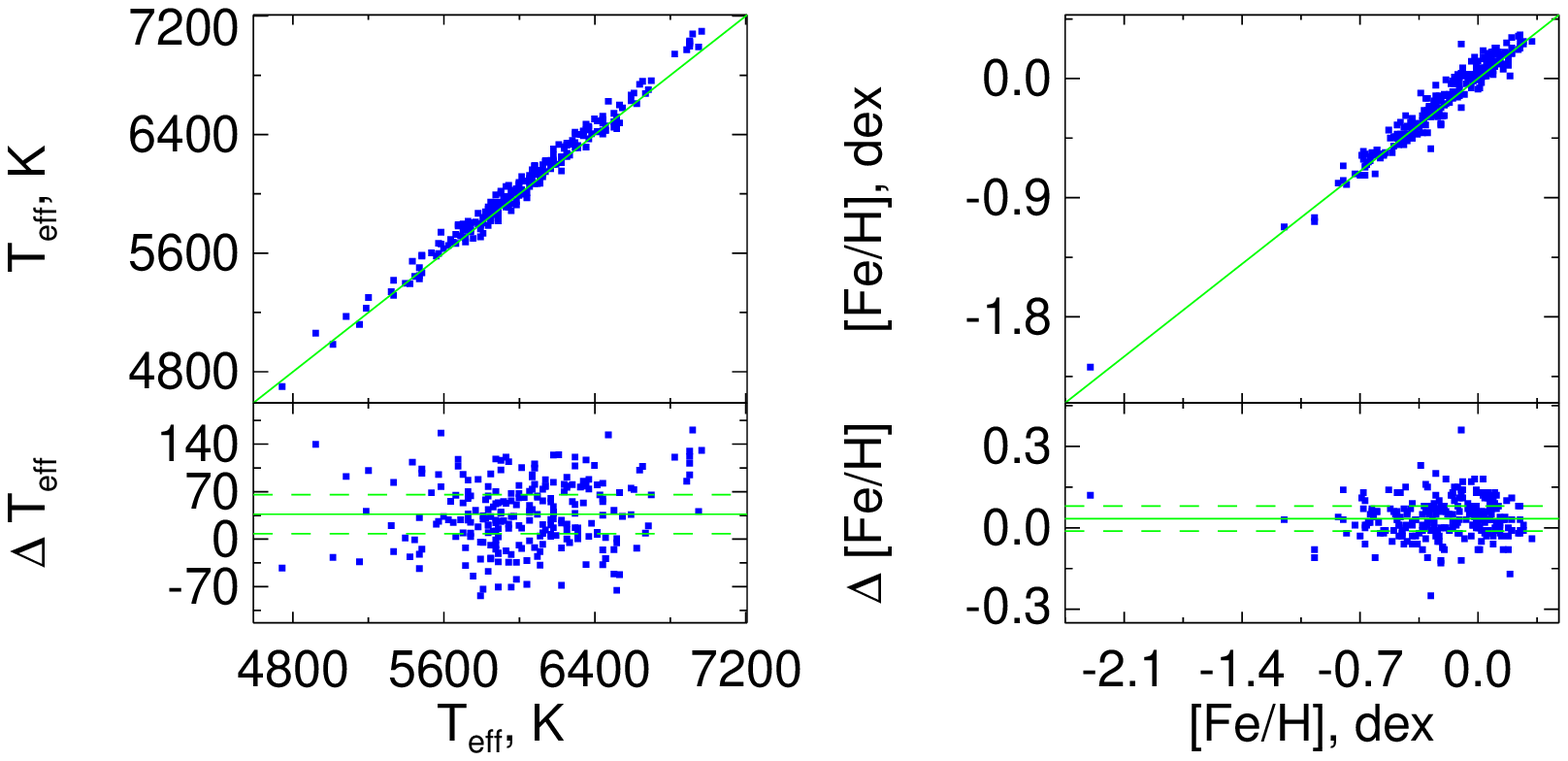}
\caption {Comparison of the atmospheric parameters determined by us
with those from \cite{nord04}. We plot 252 F, G, and K observations
in common between the two data sets, 14 $T_{\rm{eff}}$ plus two
metalicity outliers are not displayed. The axes are as in
Fig.~\ref{fig.compcflib}.}
 \label{fig.comp9}
\end{figure}

We found 269 measurements for 266 stars in common and the
corresponding comparisons are shown in Fig.~\ref{fig.comp9}, after
rejecting 15 $T_{\rm{eff}}$ outliers plus one metalicity outlier
(two measurements). One of them, HD\,72946, has been discussed in
Sec.~\ref{sec.comp12}.

For six of the outliers, HD\,54322, 124244, 137510, 37088, 94280 and
165341, our estimated parameters are consistent with the Valdes'
compilation. For HD\,82210 our value, 5343~K, is close to
\cite{mcwill90} (see Sect.~\ref{sec.comp2}). For HD\,37394, our
$T_{\rm{eff}}~=~5328$~K is 190~K hotter, and \cite{perr83} gives
5196~K from photometry. For HD\,129132, a member of a triple system,
though our determination is 200~K warmer, the quality of our fit is
satisfactory. For HD\,195633, we fitted $T_{\rm{eff}}~=~6102$~K,
\cite{nord04} give 5916~K, and  \cite{fulb00} 6000~K from
spectroscopy. For all these stars we adopt our measurements.

There is one [Fe/H] outlier HD\,22468 (K2:Vnk) with two measurements in
\citet{nord04}: [Fe/H]~=~$-1.25$~dex and $-1.52$~dex. Our
[Fe/H]~=~$-0.16$~dex is closer to the ELODIE internal determination,
[Fe/H]~=~$-0.48$~dex, and in agreement with the \ulyss{} fit to the ELODIE
spectrum: [Fe/H]~=~$-0.23$~dex $\pm 0.07$ dex. We keep our determinations.

\subsection{Comparison with \cite{kov06}}
\label{sec.comp11}

\citet{kov06} made precise temperature determinations of 215 F, G,
and K giants by measuring the line depths and equivalent widths of a
large number of spectral lines of low and high excitation potentials
and establishing $\sim$ 100 relations between $T_{\rm{eff}}$ and
their ratios. This calibration of the line depth ratio method
yielded a precision of 5 to 20~K.

\begin{figure}
\centering
\includegraphics[width=9cm]{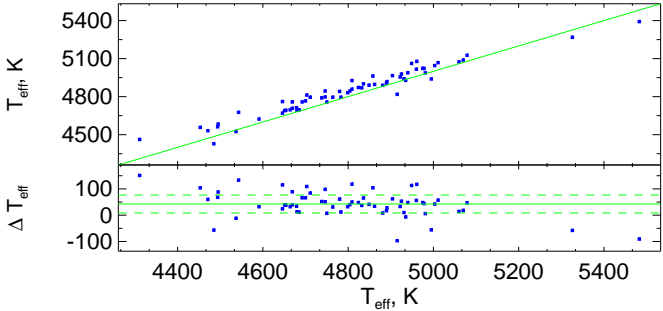}
\caption {Comparison of the atmospheric parameters determined by us
with those from \cite{kov06}. We plot 64 G and K stars in common
between the two data set, five outliers are not displayed. The axes
are as in Fig.~\ref{fig.compcflib}.}
 \label{fig.comp11}
\end{figure}

We found 69 stars in common and the comparison of $T_{\rm{eff}}$ is
shown in Fig.~\ref{fig.comp11}. There is no significant outlier.

\subsection{Comparison with \cite{mish06}}
\label{sec.comp10}

The atmospheric parameters of 177 clump giants of the Galactic disk
were determined by \cite{mish06}. The effective temperatures were
estimated by the line depth ratio method calibrated by \citet[][see
above]{kov06}. The surface gravity was estimated using the iron
ionization equilibrium and the wings of the Ca\,I triplet near
6100-6160~\AA. The [Fe/H] was derived from FeI lines. The internal
precisions are 20 K, 0.2-0.3 dex and 0.1 dex on respectively
 $T_{\rm{eff}}$, log~$g$ and [Fe/H].

\begin{figure*}
\centering
\includegraphics[width=12cm]{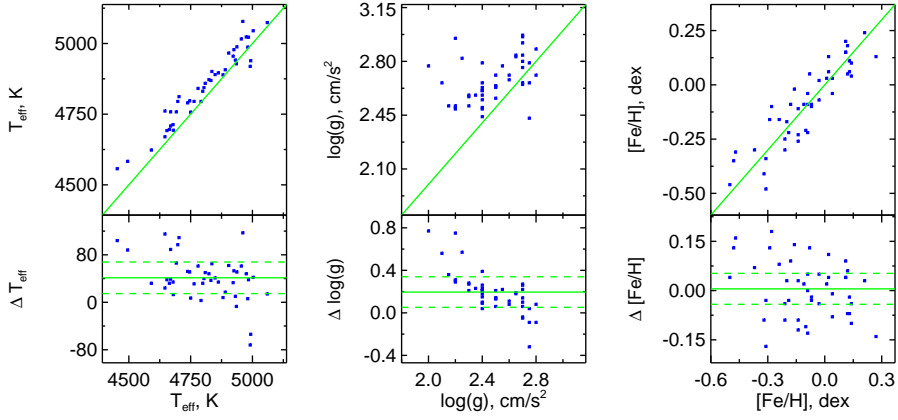}
\caption {Comparison of the atmospheric parameters determined by us
with those from \cite{mish06}. We plot 45 G and K stars in common
between the two data sets, three outliers are not displayed. The
axes are as in Fig.~\ref{fig.compcflib}.}
 \label{fig.comp10}
\end{figure*}

We identified 48 stars in common and the comparison, excluding
three $T_{\rm{eff}}$ outliers, is shown in Fig.~\ref{fig.comp10}.
These outliers were also detected in the comparison with \citet{kov06} and
are discussed in Sect.~\ref{sec.comp11}.

\subsection{Comparison with \citet{takeda07}}

\begin{figure*}
\centering
\includegraphics[width=12cm]{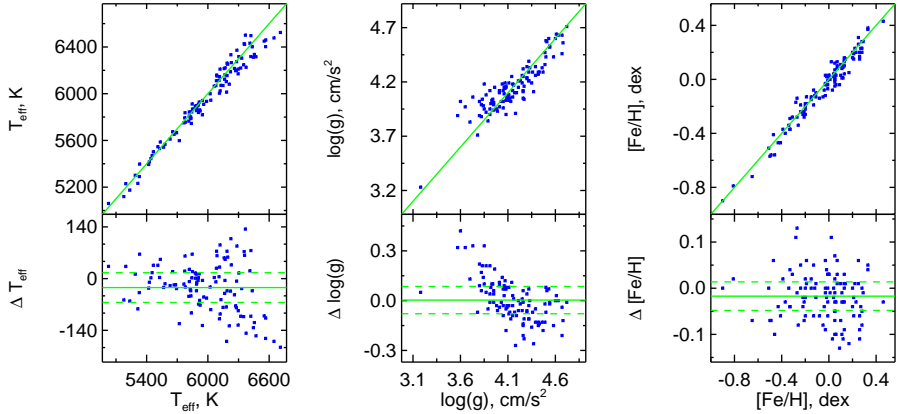}
\caption {Comparison of the atmospheric parameters determined by us
with those from \cite{takeda07}. We plot 108 F, G, and K stars in
common between the two data sets, nine outliers are not displayed.
The axes are as in Fig.~\ref{fig.compcflib}.}
 \label{fig.comp13}
\end{figure*}

We found 117 stars in common with \citet{takeda07}, who estimated
the atmospheric parameters of solar analogs from the analysis of
R=70\,000 spectra. The internal precisions of this series of
measurements are 17~K, 0.04 and 0.02 dex on respectively
$T_{\rm{eff}}$, log$g$ and [Fe/H]. The comparison is shown in
Fig.~\ref{fig.comp13}. Nine $T_{\rm{eff}}$ outliers were clipped
from the statistics given in Table \ref{tb.fgk}. For eight of them
HD\,121370, 99747, 98991, 151769, 137510, 128167, 15798, 4307, our
parameter estimations are quite close to the Valdes adopted values.
For HD\,224930 our temperature (5427~K) agrees with that of
\cite{soub08} (5413~K) and ELODIE (5446~K), while \citet{takeda07}
give 5681~K and \cite{fulb00} 5275~K. We maintain our original
measurements.

\subsection{Comparison with \cite{Hekker07}}

\cite{Hekker07} used high S/N spectra at R=60\,000 of 380 G and K
giant stars to measure the atmospheric parameters. The effective
temperatures, surface gravities, and metalicities are determined
from the equivalent width of iron lines, by imposing excitation and
ionization equilibria. They compared their stellar parameters with
\citet{ram05}, an updated version of the \citet{cayrel01}
compilation, for 254 stars in common. They found their effective
temperature and surface gravity are higher by 56~K and 0.15~dex, and
with 84~K and 0.22~dex dispersion respectively. The dispersion of
[Fe/H] is 0.10~dex. They found their results to be consistent with
\citet{luck07}.

We compare the 132 stars in common after clipping five outliers
(Fig.~\ref{fig.comp13}). For HD\,39118 (G8III+...), our
$T_{\rm{eff}}~=~4927$~K is 377~K hotter while the ELODIE internal
temperature is 5029~K, and the \ulyss{} fit to the ELODIE spectrum
gives 5000~K. For HD\,184406 and 29139 our estimates are consistent
with both the Valdes compilation and with the ELODIE internal
determinations. The temperatures given by \citet{Hekker07} are
$\sim$240~K hotter. For HD\,115004 and 57669, our temperatures are
$\sim$170~K hotter, but we have no reason to question our fit. In
all cases we adopt our solutions.

\begin{figure*}
\centering
\includegraphics[width=12cm]{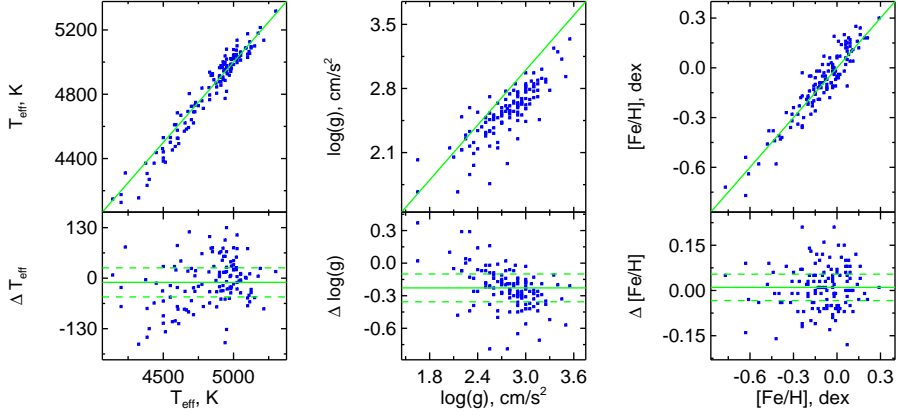}
\caption {Comparison of the atmospheric parameters determined by us
with those from \cite{Hekker07}. We plot 127 G and K stars in common
between the two data sets, five outliers are not displayed. The axes
are as in Fig.~\ref{fig.compcflib}.}
 \label{fig.comp14}
\end{figure*}

\subsection{Comparison with \cite{luck07}}

\begin{figure*}
\centering
\includegraphics[width=12cm]{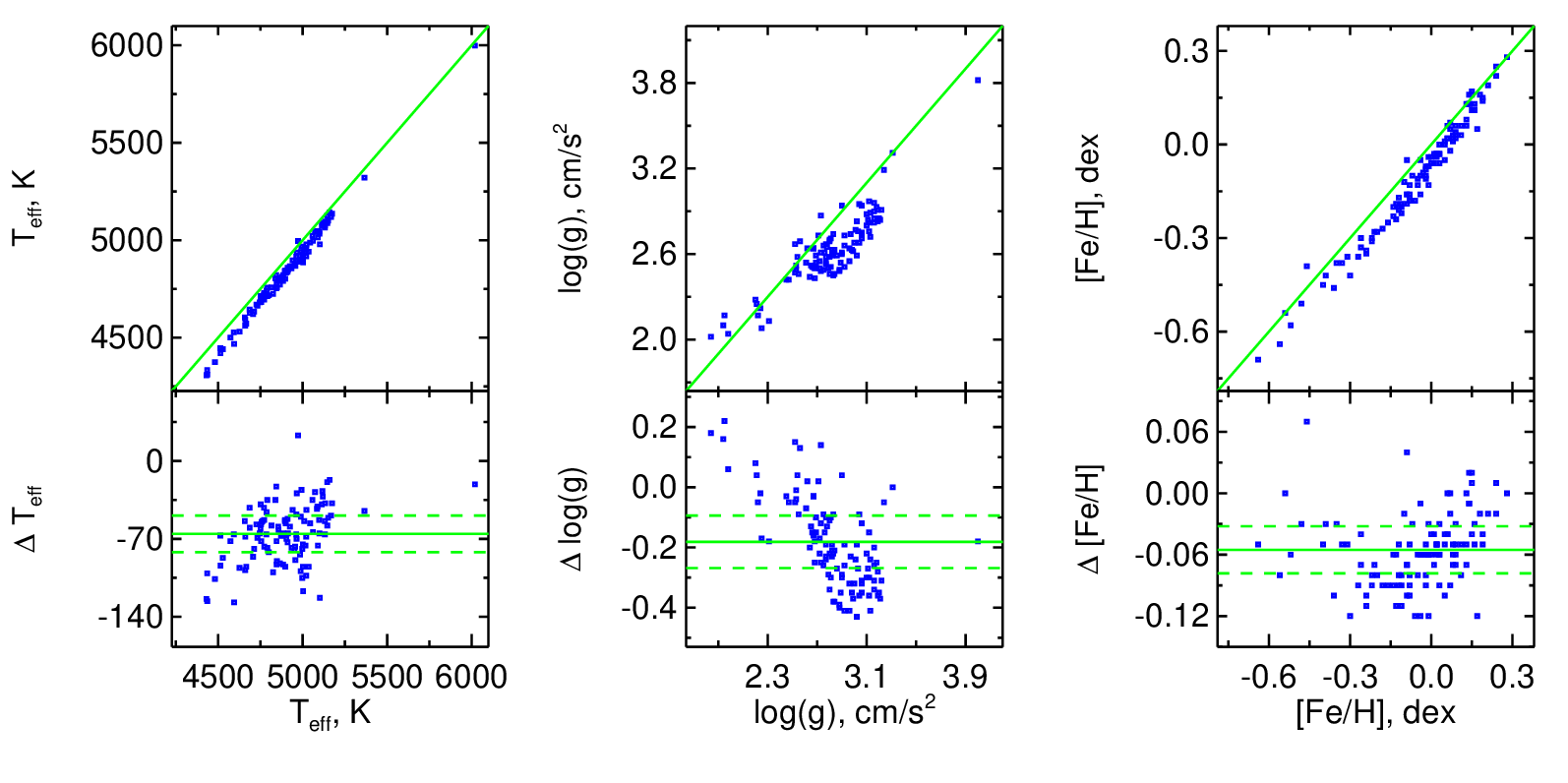}
\caption {Comparison of the atmospheric parameters determined by us
with those from \cite{luck07}. We plot 108 G and K stars in common
between the two data sets, five outliers are not displayed. The axes
are as in Fig.~\ref{fig.compcflib}.}
 \label{fig.comp15}
\end{figure*}

\cite{luck07} present the parameters for 298 nearby G and K giants
using spectroscopy and photometry. The external uncertainty of their
temperatures is on the order of $\sim$100~K. We compared the 113
stars found in common (Fig.~\ref{fig.comp15}) and found a mean
difference in temperature of -66~K (\ulyss{} minus \cite{luck07};
see Table~\ref{tb.fgk}).

There are a total of five $T_{\rm{eff}}$ outliers, HD\,102328,
85503, 104979, 167768 and 126271, our temperatures are between
150-230~K lower. HD\,85503 is already discussed in
Sect.~\ref{sec.comp11}. For HD\,104979, our result (4818~K) agrees
well with the ELODIE internal determination (4814~K) and with the
Valdes compilation (4850~K), while \cite{luck07} give 4996~K.

\subsection{Comparison with \cite{sousa08}}
\label{sec.comp4}

\begin{figure*}
\centering
\includegraphics[width=12cm]{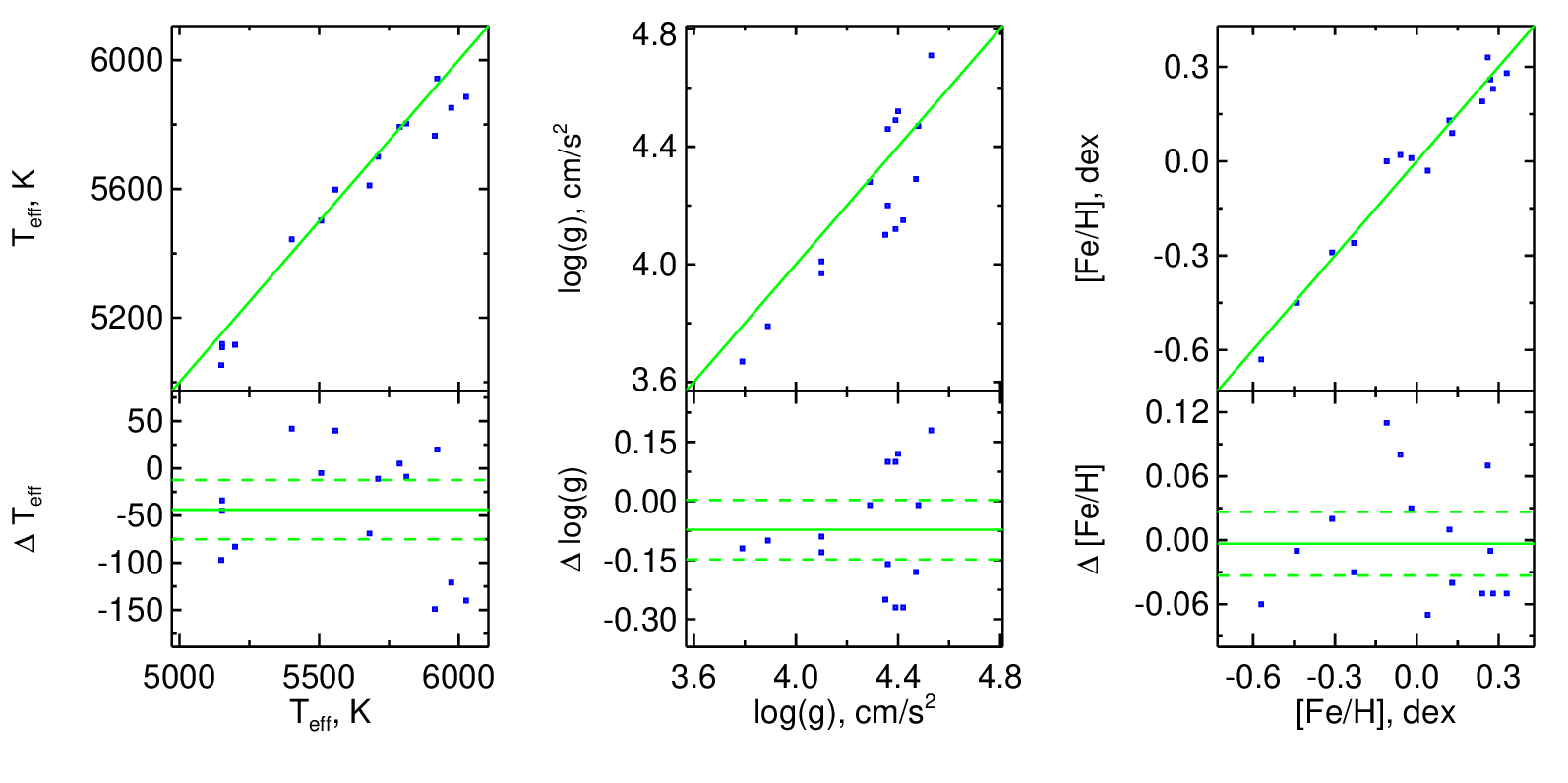}
\caption {Comparison of the atmospheric parameters determined by us
with those from \cite{sousa08}. We plot 17 F, G, and K stars in
common between the two data sets, two outliers are not displayed.
The axes are as in Fig.~\ref{fig.compcflib}.}
 \label{fig.comp16}
\end{figure*}

\cite{sousa08} give accurate stellar parameters for 451 stars using
high-resolution spectra. We compared the 17 common FGK observations
on Fig.~\ref{fig.comp16} where 2 $T_{\rm{eff}}$ outliers have been
clipped.

HD\,19994 (F8V) was already discussed in Sec.~\ref{sec.comp4},
and the other outlier is not significant.

\section[]{External comparisons for OBA stars} \label{appendix:comparisonoba}

In this appendix we present detailed comparisons between our
measurements and four previous studies.

\subsection{Comparison between ELODIE 3.2 absolute and internal parameters}
The absolute and internal atmospheric parameters of ELODIE are
described in Sect.~\ref{sect:elo32}. We stress that the present
comparison is different from the one in Sect.~\ref{sec.consistency},
where we used \ulyss{} to determine the atmospheric parameters of
the ELODIE observations.

\begin{figure*}
\centering
\includegraphics[width=12cm]{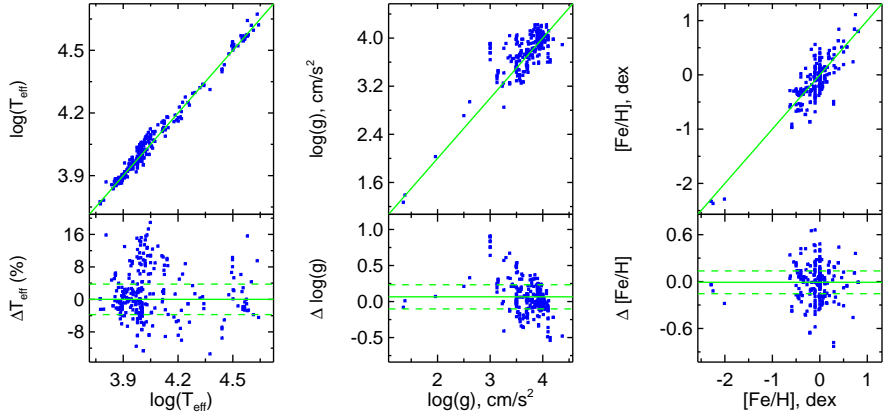}
\caption {Comparison of the ELODIE 3.2 Absolute (A) and Internal (I)
atmospheric parameters. We plot 274 O, B, and A type stars, 18
$T_{\rm{eff}}$ outliers plus one metallicity outlier are not
displayed. The axes are similar as in Fig.~\ref{fig.compcflib}, all
the abscissas are the measurements of A. $T_{\rm{eff}}$ is in log10.
$Upper$ panels: the ordinates are from I; $Lower$ panels: the
ordinates are the difference I $-$ A.} \label{fig.compoba4}
\end{figure*}

There are 293 O, B, and A type stars with valid internal
determinations in ELODIE stellar library. The comparison between the
absolute and internal values of the parameters are shown in
Fig.~\ref{fig.compoba4} after excluding 18 $T_{\rm{eff}}$ outliers
and one metalicity outlier. The comparison statistics are listed in
Table \ref{tb.oba} without these 19 outliers.

The absolute parameters of almost all of these outliers were not
available from spectroscopic analyses in the literature. The
temperature were estimated from the  B-V colur (Tycho-2 catalog of
\citealt{hog00}), assuming an empirical color-temperature relation
for a main-sequence star; log~$g$ were converted from the V absolute
magnitude from Hipparcos and $T_{\rm{eff}}$, using a bolometric
correction valid for a main-sequence star and an empirical
mass-to-light relation. See \cite{PS01} for more details.

We did not find any significant bias on the determination of the
three atmospheric parameters, but the dispersions are higher than
for the FGK stars. This is primarily because of the lack of accurate
measurements for the reference stars.

In addition, the modeling of the atmosphere with the three
fundamental parameters is an over-simplification, because the spread
of other physical characteristics, which we neglected, induces a
dispersion of our measurements. The modeling can probably be
improved in the future, but the important point is that the present
method apparently does not introduce major biases.

\subsection{Comparison with ELODIE 3.2}
\label{sec.compoba1} 
\begin{figure*}
\centering
\includegraphics[width=12cm]{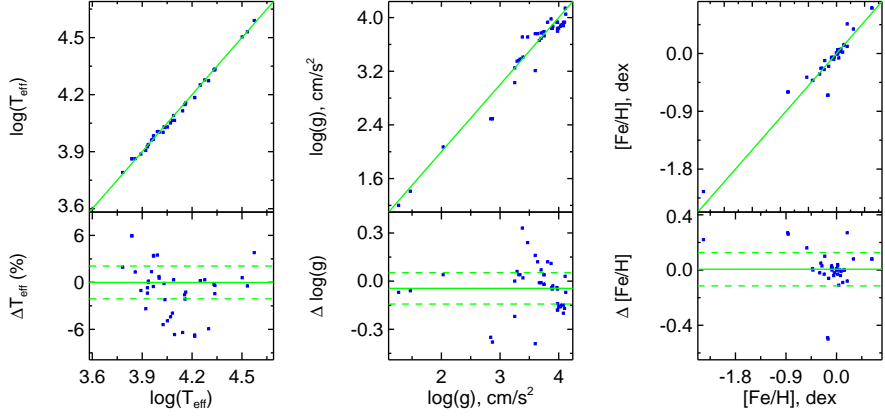}
\caption {Comparison of the atmospheric parameters determined by us
with those from ELODIE 3.2. We plot 41 O, B, and A type stars in
common between the two data sets, six outliers are not displayed.
The axes are as in Fig.~\ref{fig.compcflib}, besides $T_{\rm{eff}}$
is in log10.} \label{fig.compoba1}
\end{figure*}

There are 47 O, B, and A observations in common
with ELODIE. The comparison with the ELODIE internal determinations
is shown in Fig.~\ref{fig.compoba1} after excluding six
$T_{\rm{eff}}$ outliers.

For the first outlier, HD\,212571 (B1Ve), though it is a Be type
star, we could not see any emission lines in both its CFLIB and
ELODIE observations, but the spectrum is quite rotationally
broadened. Its ELODIE internal parameters are 20530~K, 3.29~dex,
0.01~dex, and we obtain 23278~K, 3.53~dex, -0.02~dex, which are
close to the ELODIE absolute parameters, 23714~K, 3.50~dex and
0.00~dex. Fitting the ELODIE spectrum, we get 23031~K, 3.41~dex and
-0.06~dex, in agreement with our measurements for CFLIB. We suppose
that the discrepancy results from the different approach for
modeling the effect of stellar rotation in our \ulyss{} measurements
and in the internal ELODIE measurement (see
Sect.~\ref{sec.consistency}).

The second outlier, HD\,5394 (B0IVpe), displays prominent $\rm
H_{\alpha}$ and $\rm H_{\beta}$ emission lines (rejected when
performing the \ulyss{} fit). There are two observations of this
star in the ELODIE library, with determinations: [34097~K, 3.34~dex,
0.11~dex] and [39156~K, 3.22~dex, -1.56~dex]. Because of the
emission lines, these spectra were not used to build the ELODIE
interpolator. Our results [33939~K, 3.38~dex, 0.02~dex] are close to
the former ELODIE internal values. The \ulyss{} fit of these ELODIE
spectra provides 33090~K, 3.41~dex and 0.05~dex, consistent with our
measurements.

For the third outlier HD\,206165 (B2Ib), the ELODIE internal
determinations are 19685~K, 2.94~dex and -0.27~dex, while we get
22112~K, 3.09~dex and -0.12~dex. \cite{doug92} derived 19040~K,
2.61~dex and -0.33~dex comparing Str\"{o}mgren de-reddened color
indices and H$\gamma$ line profiles to line-blanketed atmosphere
models \citep{kurucz79}. The abundances were derived using LTE
Kurucz models. They quote errors of  2-4\% on $T_{\rm{eff}}$ and 0.1
dex on log~$g$. So the ELODIE internal solution is compatible with
their measurements. The \ulyss{} fit of the ELODIE spectrum gives
20569~K, 2.93~dex and -0.28~dex, also consistent with the ELODIE
internal determinations. We adopt this latter solution rather than
our fit of the CFLIB spectrum.

The fourth outlier is HD\,86986 (A1V), a blue horizontal branch star
(BHB). Our determinations, 8843~K, 4.28~dex and -0.76~dex, are 676~K
warmer than the ELODIE internal value. \citet{kinman00} and
\citet{behr03} give 7950~K and 7775~K respectively. The ELODIE
interpolator is only weakly constrained in this region of the
parameters' space, and we adopt the average of the two recent
literature measurements.

For the fifth outlier, HD\,176437 (B9III), the ELODIE internal
parameters are 12230~K, 4.08~dex and 0.25~dex, and we find 11163~K,
4.06~dex and 0.15~dex. \cite{bala86} give $T_{\rm{eff}}$=10080~K.
The last outlier, HD\,220825 (A0p...), is a CVn star that is
discussed below in Sect.~\ref{sec.compoba2}. The ELODIE internal
yield is 9683~K, 3.80~dex and 0.47~dex, while we find 10375~K,
3.78~dex and 0.74~dex. This is 692~K warmer and is consistent with
\cite{leon66}: 10286~K. For these two stars, we adopt our
measurements.

\subsection{Comparison with \cite{cflib}}
\label{sec.compoba2} 
\begin{figure*}
\centering
\includegraphics[width=12cm]{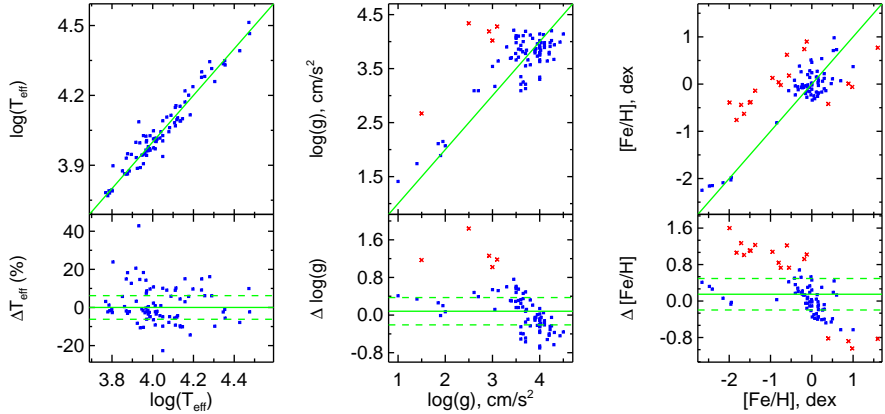}
\caption {Comparison of the atmospheric parameters determined by us
with those from \cite{cflib}. We plot 84 O, B, and A type stars in
common between the two data sets. Two $T_{\rm{eff}}$ outliers are
not displayed. Five log~$g$ and 18 [Fe/H] discrepant measurements
are shown as red crosses and are discussed in the text. The axes are as in
Fig.~\ref{fig.compcflib}, besides $T_{\rm{eff}}$ is in log10.}
\label{fig.compoba2}
\end{figure*}

Of the 260 O, B, and A stars in CFLIB, the
atmospheric parameters were compiled for only 86 stars by
\cite{cflib}. The comparison between these values and our
determinations is displayed in Fig.~\ref{fig.compoba2} after
excluding two temperature outliers. The 18 cases where the [Fe/H]
deviations between CFLIB and our measurements  is greater than
0.7~dex are displayed in red (right panel), as well as the five
log~$g$ outliers deviating by more than 0.8~dex (central panel).
These five stars, HD\,105262, 161817, 18296, 2857 and 86986, have
also discrepant metalicities. All these cases are discussed below.

The statistics, with and without the outliers,
are reported in Table~\ref{tb.oba}.

For HD\,4727 (B5) and HD\,39283 (A2V), the large discrepancy on
$T_{\rm{eff}}$ is due to confusions on the designations in the Valdes
compilation. Our measurements are consistent with the spectral
classification.

Four stars are classified CVn, chemically peculiar stars, HD\,18296,
34797, 72968 and 220825, for which Valdes adopted low metalicities
taken from old curve-of-growth references. More recently,
\citet{alonso03} studied HD\,34797 and found a significant
over-abundance. This supports our solution ([Fe/H]~=~0.62~dex). For
HD\,220825, \cite{glago06} give
[Fe/H]~=~0.69~dex, consistent with our result and with the
measurements on the ELODIE spectrum. We did not find any detailed
analysis on the metalicity of the two last CVn, HD\,18296 and 72968.
\citet{leone98} and \citet{takeda09} studied another supposedly CVn
star, HD79469, and re-qualified it as a `normal' star, although we
find a significant over-abundance. These comparisons tend to grant
confidence in the ability of our method to retrieve the metalicity
of the chemically peculiar stars, and we therefore adopt our
solutions.

Among the [Fe/H] outliers are five BHB stars HD\,2857, 74721, 86986,
109995 and 161817, and one post-AGB HD\,105262. Their atmospheric
parameters determined by \citet{kinman00} and \citet{behr03} are
consistent with the previous measurements adopted in Valdes, but
more precise and reliable. As pointed out in
Sect.~\ref{sec.compoba1}, the ELODIE interpolator is not performing
well in this region. For these six stars, we obtained a mean [Fe/H]
of $-0.46$~dex, while the average of the aforementioned references
is $-1.64$~dex. We adopt averages of these references rather than
our own measurements.

For HD\,27295, we give [Fe/H]~=~$-0.02$~dex and Valdes take $-0.75$~dex,
from \citet[][analysis of IUE spectra]{smith93}.
\citet{behr03} obtained recently [Fe/H]~=~$-0.95\pm0.06$ and
[Mg/H]~=~$-0.46\pm0.05$ for this main sequence chemically peculiar star.
We adopt the \citet{behr03} solution.

For HD\,183324, Valdes quotes -1.50~dex from \cite{sturen93}.
\ulyss{} returned a value of [Fe/H]~=~$-0.39$~dex.
Recently, \citet{saffe08} determined [Fe/H]~=~$-1.22$$\pm0.30$~dex,
and we adopt their solution.

The metalicity reported in Valdes for HD\,60179 \citep{smith74}
appears to be erroneous (transcription or conversion error). The
star has a solar metalicity, and there is no disagreement with our
estimate. For HD174959, \citet{heacox79} gives [Fe/H]~=~$-0.8$
(adopted by Valdes), but also [Mg/H]~=~$-0.3$ and a solar abundance
of Ni. So, the low [Fe/H] may not be real. There is no other
detailed analysis of this star. For HD\,155763, we find
[Fe/H]~=~0.13~dex, while Valdes report $-0.95$~dex. More recently,
\citet{adel98} provides $T_{\rm{eff}}$~=~12500, log~$g$~=~3.50 and
[Fe/H]~=~$-0.11$, from R~$\sim$~50\,000 optical spectra. For
HD\,175640, a typical HgMn star, we give [Fe/H]~=~0.18~dex and
Valdes adopt -0.55~dex from \citet{smith93}. \citet{castelli04} give
[Fe/H]~=~$-0.02\pm0.10$, from high quality, R~$\sim$~100\,000,
spectra. For HD\,58343 (B2Vne), we estimate [Fe/H]~=~0.01~dex, while
Valdes quotes 0.89~dex from \cite{koda70}. \citet{fremat05}
determined the temperature and the surface gravity, confirming that
the temperature used by \cite{koda70} is too hot. This may be an
explanation for their high metalicity. For these five stars, we
adopt our original measurements.

We conclude that the trends seen in Fig.~\ref{fig.compoba2} are
partly caused by an over-estimate of the metalicity of the
metal-poor evolved stars by our analysis, and partly by inaccurate
measurements in the literature.

\subsection{Comparison with \cite{cena07}}
\begin{figure*}
\centering
\includegraphics[width=12cm]{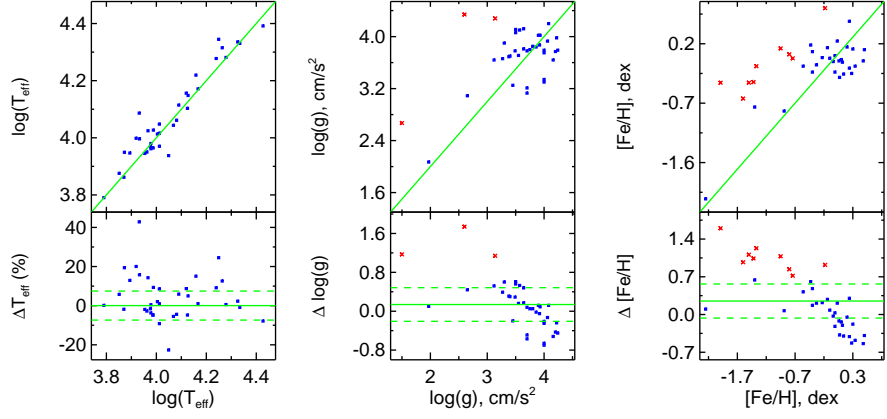}
\caption {Comparison of the atmospheric parameters determined by
us with those from \cite{cena07}. We plot 37 B and A type stars
in common between the two data sets. One $T_{\rm{eff}}$ outlier is
not displayed. Three log~$g$ and nine [Fe/H] deviating measurements
are displayed as red crosses and discussed in the text.
The axes are as in Fig.~\ref{fig.compcflib},
besides $T_{\rm{eff}}$ is in log10.} \label{fig.compoba3}
\end{figure*}

\cite{cena07} compiled and homogenized the atmospheric parameters
for MILES \citep{miles}. This library contains 985 stars, spanning a
large range in atmospheric parameters. There are 38 common B and A
type stars between CFLIB and MILES, and we show the comparison in
Fig.~\ref{fig.compoba3} after excluding one $T_{\rm{eff}}$ outlier.
Several log~$g$ and [Fe/H] outliers are shown by red crosses. The
comparison statistics listed in Table~\ref{tb.oba} excludes the one
$T_{\rm{eff}}$ outlier and all the log~$g$ and [Fe/H] outliers.

The $T_{\rm{eff}}$ outlier is HD\,89822, for which Valdes adopts
10500~K 
from \cite{smith93},
consistent with our estimates of 10286~K.
For this star, \cite{cena07} give an erroneous 5538~K.

Besides this detected $T_{\rm{eff}}$ outlier, a couple of other
cases catch the attention in Fig.~\ref{fig.compoba3} (left panel).
The most prominent is HD\,105262 (B9). \cite{cena07} and Valdes
adopted the same set of parameters, and the star is already
discussed in Sect.~\ref{sec.compoba2}. The second one is HD\,206165,
already discussed in Sect.~\ref{sec.compoba1}.

We find also nine [Fe/H] outliers displayed in red in
Fig.~\ref{fig.compoba3} (right panel). These are HD\,2857, 27295,
74721, 105262, 109995, 155763, 174959, 183324, and 220825. Finally,
there are three log~$g$ cases displayed in red in
Fig.~\ref{fig.compoba3} (middle panel): HD\,2857, 86986, and 105262.
All these stars are already discussed in Sect.~\ref{sec.compoba2}.

 As for Fig.~\ref{fig.compoba2}, the trends seen in
Fig.~\ref{fig.compoba3} result from a combination between wrong
measurements reported in the literature and over-estimated
metalicities for the metal-poor evolved stars.

\section[]{External comparisons for M stars} \label{appendix:comparisonm}

In this appendix we present detailed comparisons between our
measurements and four previous studies.

\subsection{Comparison with ELODIE 3.2}
\begin{figure*}
\centering
\includegraphics[width=12cm]{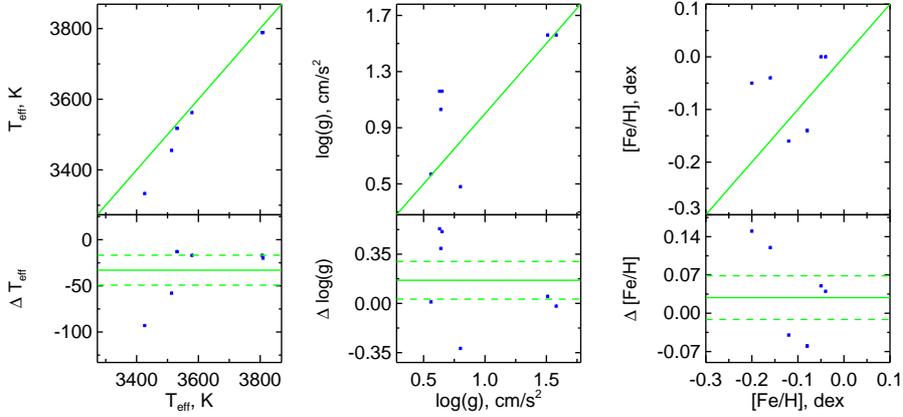}
\caption {Comparison of the atmospheric parameters determined by us
with those from ELODIE 3.2 for the seven M type stars in common. The
axes are as in Fig.~\ref{fig.compcflib}.}
 \label{fig.compm1}
\end{figure*}

The first comparison is with the ELODIE (version 3.2) internal
parameters. There are six M and one S type common observations for
five stars between the ELODIE and CFLIB library. The comparison is
shown in Fig.~\ref{fig.compm1}.

The most $T_{\rm{eff}}$ departing star is HD\,175588. ELODIE
internal temperature is 3426~K, while we give 3333~K.

\subsection{Comparison with \cite{cflib}}
\begin{figure*}
\centering
\includegraphics[width=12cm]{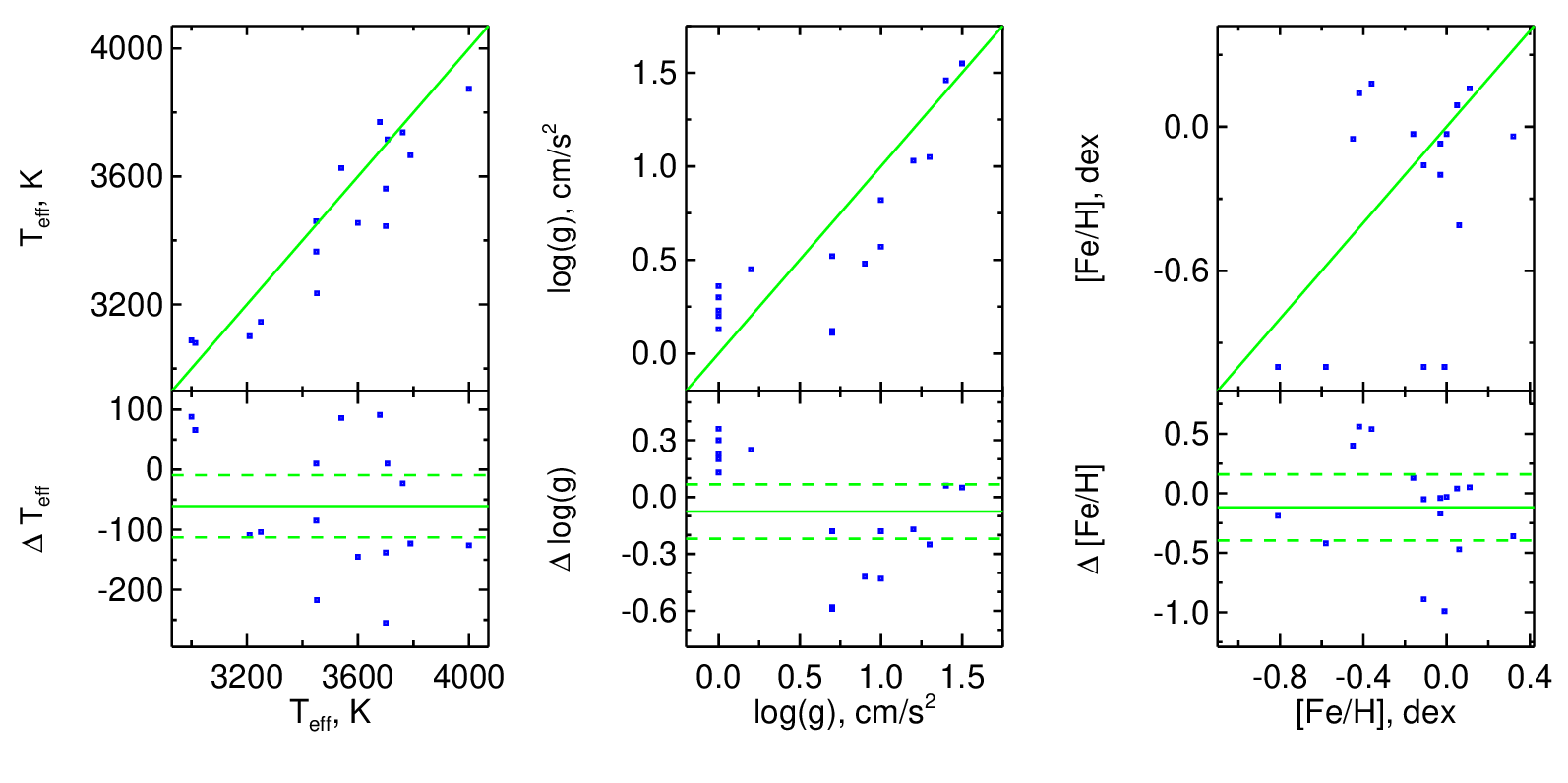}
\caption {Comparison of the atmospheric parameters determined by us
with those from \cite{cflib}. We plot 16 M type stars in common
between the two data sets, two outliers are not displayed. The axes
are as in Fig.~\ref{fig.compcflib}.}
 \label{fig.compm2}
\end{figure*}

The comparison with the parameters compiled in \cite{cflib} for 17 M
and 1 S type stars is shown in Fig.~\ref{fig.compm2} after excluding
two $T_{\rm{eff}}$ outliers. For the first, G\_103-68 (M3), Valdes
quote a temperature of 5511~K from \cite{carn94} based on
photometry, while we give 3477~K, which corresponds better to the
spectral type. For the other star G\_176-11 (M2V), Valdes report
3544~K from \cite{worthey94}, while we determine 3948~K, 5.13~dex
and -1.43~dex for $T_{\rm{eff}}$, log~$g$ and [Fe/H] respectively.
\cite{soub08} give 3687~K, 4.90~dex and -0.43~dex. We adopt our
solution.

\subsection{Comparison with \cite{cena07}}
\begin{figure}
\includegraphics[width=9cm]{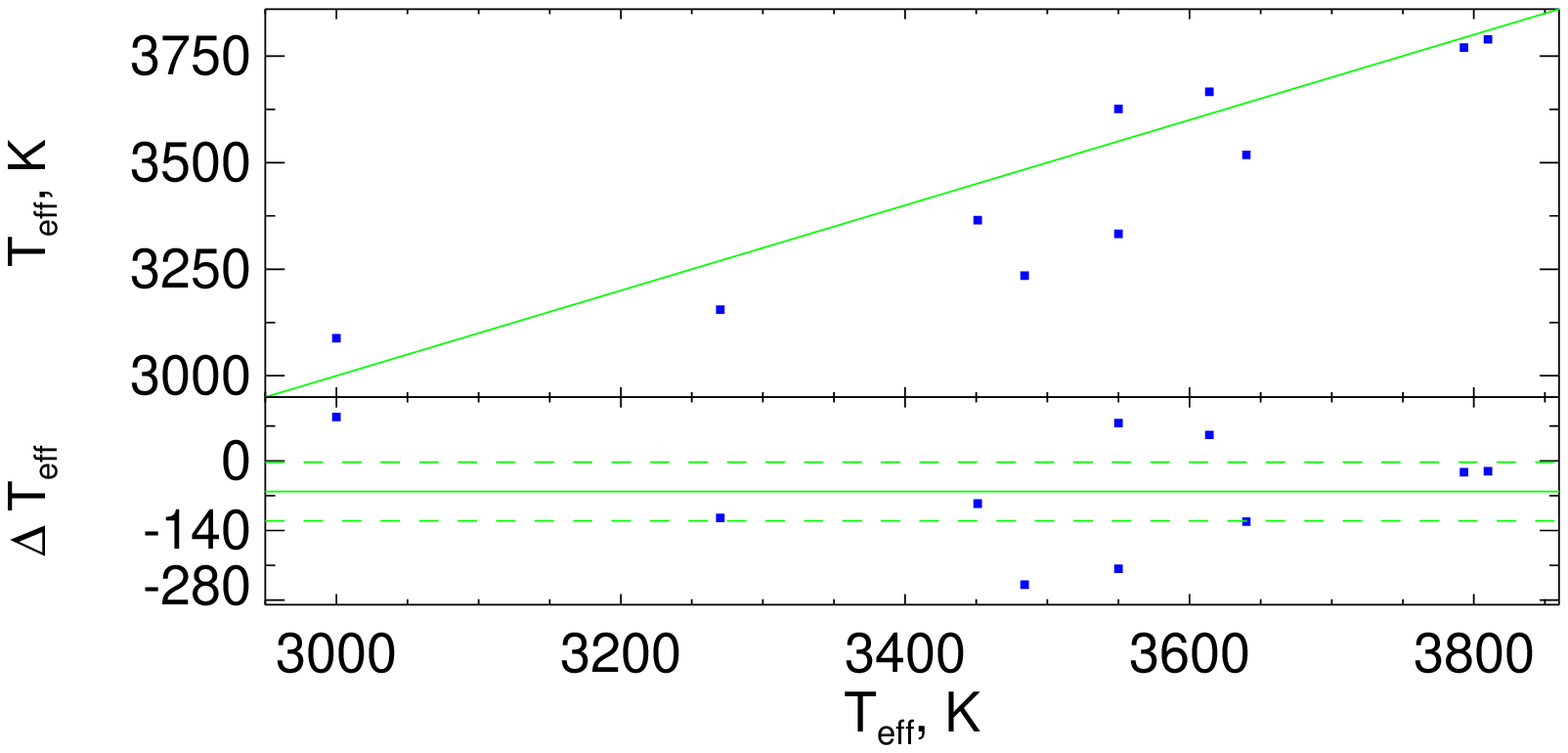}
\caption {Comparison of the atmospheric parameters determined by \ulyss{}
with those from \cite{cena07} for 10 M type stars in common. The axes are
as in Fig.~\ref{fig.compcflib}.}
 \label{fig.compm3}
\end{figure}

We found ten M type stars in common with \cite{cena07}, but three of them
lack log~$g$ and half of them miss [Fe/H] values in \cite{cena07}.
Therefore, we only compare the effective temperatures between our d
determinations and this reference, the result is shown in
Fig.~\ref{fig.compm3}. The most deviation with 249~K cooler is star
HD\,123657, \cite{cena07} give 3484~K, log~$g$\,=\,0.85~dex, \ulyss{}
fitted with 3235~K, 0.48~dex and -0.20~dex. CFLIB adopt 3452~K,
0.90~dex and -0.03~dex \cite{smith85}.

\subsection{Comparison with \cite{cayrel01}}
\begin{figure*}
\centering
\includegraphics[width=12cm]{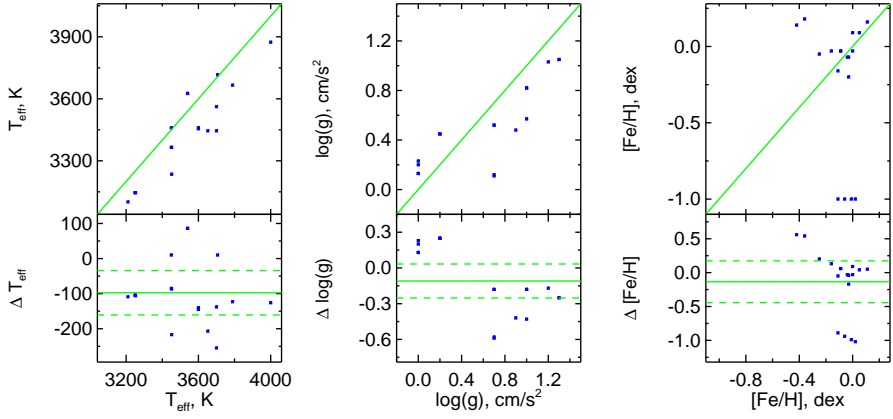}
\caption {Comparison of the atmospheric parameters determined by
\ulyss{} with those from \cite{cayrel01}. We plot 18 M type
observations in common between the two data sets, actually 12
stars. The axes are as in Fig.~\ref{fig.compcflib}.}
 \label{fig.compm4}
\end{figure*}

 \cite{cayrel01}  present a compilation
of published atmospheric parameters obtained from various sources
and hence their data are inhomogeneous. The comparison
between their parameters and ours is shown in
Fig.~\ref{fig.compm4}. We identified 11 M and 1 S type stars in common,
for a total of 18 measurements in this compilation.

\end{appendix}

\setcounter{table}{3}  
\onltab{3}{
\longtab{3}{
\setlength{\tabcolsep}{2.5pt}
\begin{longtable}{l|rrr|rrr|l|rrr|c}
\caption{\label{tb.outlier}79 Outliers found in the comparisons with the literature.}\\
\hline\hline
Name & $T_{\rm eff}$ & $log(g)$ & [Fe/H] & $T_{\rm eff}$ & $log(g)$ & [Fe/H] & Reference & $T_{\rm eff}$ & $log(g)$ & [Fe/H] & Flag \\
     & (K)         & ($cm\,s^{-2}$) & (dex) & (K)       & ($cm\,s^{-2}$)  & (dex) &        & (K)         & ($cm\,s^{-2}$) & (dex) &    \\
\hline

\endfirsthead
\caption{continued.}\\
\hline\hline

Name & $T_{\rm eff}$ & $log(g)$ & [Fe/H] & $T_{\rm eff}$ & $log(g)$ & [Fe/H] & Reference & $T_{\rm eff}$ & $log(g)$ & [Fe/H] & Flag \\
     & (K)         & ($cm\,s^{-2}$) & (dex) & (K)       & ($cm\,s^{-2}$)  & (dex) &        & (K)         & ($cm\,s^{-2}$) & (dex) &    \\
\hline
\endhead
\hline
\endfoot

HD002857  &  8895 & 4.34 & -0.63 &  7522 & 2.50 & -1.64 & \cite{kloch87}  &  7776 & 3.19 & -1.70 & 2\\
          & (101) &(0.07)& (0.16)&  7550 & 3.00 & -1.73 & \citet{kinman00}&       &      &       &  \\
          &       &      &       &  8002 & 3.38 & -1.67 & \citet{behr03}  &       &      &       &  \\
HD004307  &  5803 & 4.01 & -0.26 &  5648 & 3.75 & -0.29 & \cite{takeda07} &  5803 & 4.01 & -0.26 & 0\\
          &  (32) &(0.06)& (0.05)&  5809 & 4.06 & -0.28 & \cite{ed93}     &       &      &       &  \\
HD005394  & 33939 & 3.38 &  0.02 & 33090 & 3.41 &  0.05 & fit\_ELO        & 33939 & 3.38 &  0.02 & 0\\
          &(1090) &(0.06)& (0.04)& 34097 & 3.34 &  0.11 & ELODIE\_I       &       &      &       &  \\
          &       &      &       & 39156 & 3.22 & -1.56 & ELODIE\_I       &       &      &       &  \\
\hline
\end{longtable}
\tablefoot{ The first column is the name of the outlier star (see
Sect.~\ref{sec.fgk}). In the second to fourth columns, the first
line is our fitted $T_{\rm eff}$, $log(g)$ and [Fe/H], the second
line is the internal errors. Columns five to seven are the
parameters published by other references. Column eight lists the
reference, where `ELODIE\_I' means ELODIE internal determinations,
`fit\_ELO' means \ulyss{} fit of an ELODIE spectrum, `fit\_3900'
means fit of the CFLIB spectrum starting from 3900~\AA{} instead of
4400~\AA. Columns nine to eleven are the adopted values in our final
table. The last column is a flag describing the adopted solution:
`0', standard fit of the CFLIB spectrum; `1', fit of the CFLIB
spectrum starting from 3900~\AA; `2', from literature; `3', \ulyss{}
fit of ELODIE spectra; `4', improved fit, see the text.
} 
} 
} 

\end{document}